\documentclass[%
 amsmath,amssymb,
]{revtex4-2}

\usepackage{dcolumn} % Align table columns on decimal point       
\usepackage{bm}      % bold math
\usepackage{setspace}
\usepackage{amsfonts}
\usepackage{braket}

\usepackage{graphicx}% Include figure files
\usepackage{amsmath}	% required for `\align' (yatex added)
\begin{document}

%\preprint{APS/123-QED}

\title{100 Mfps ghost imaging with wavelength division multiplexing}

\author{
Shin Motooka$^{1}$, 
Noriki Komori$^{1}$, 
Tomoaki Niiyama$^{2}$, 
and 
Satoshi Sunada$^{2}$
}

\affiliation{${}^{1}$Graduate School of Natural Science and Technology, Kanazawa University, Kakuma-machi, Kanazawa, Ishikawa, 920-1192, Japan\\ ${}^{2}$Faculty of Mechanical Engineering, Institute of Science and Engineering, Kanazawa University, Kakuma-machi, Kanazawa, Ishikawa, 920-1192, Japan}


\begin{abstract} 
Ghost imaging (GI) and single-pixel imaging (SPI) techniques enable image reconstruction without spatially resolved detectors, offering unique access to wide spectral ranges and challenging imaging environments. Yet, their adoption has been limited by the slow generation of mask patterns, which constrains achievable frame rates. Here, we demonstrate ultrafast GI that achieves a spatial-temporal information flux of 78.4 gigapixels per second across five wavelengths, which is at least two orders of magnitude larger than that reported for previous training-data-free GI approaches. This breakthrough is enabled by 25 GHz speckle pattern switching and allows parallelizing the pattern illumination using a wavelength-division multiplexing (WDM) technique. We show that the proposed approach is capable of reconstructing 28$\times$28-pixel images at the exposure time of 10 ns, achieving 100 megaframes per second (Mfps), and demonstrate the GI of a microsecond-scale dynamic event. This approach opens avenues for studying rapid processes in physics, chemistry, and biology, where conventional cameras are limited by detector bandwidth, readout speed, or cost.
\end{abstract}

%%%%%%%%%%%%%%%%%%%%%%%%%%  body  %%%%%%%%%%%%%%%%%%%%%%%%%%
\maketitle
\section{Introduction}
Single-pixel and ghost imaging (SPI/GI) have emerged as powerful computational imaging techniques capable of reconstructing images from total intensity measurements under sequentially projected spatial patterns. Unlike conventional cameras, SPI/GI requires only a single photodetector, making it particularly suitable for spectral regimes where standard imaging sensors are inefficient, as well as under low-light or otherwise challenging experimental conditions \cite{Edgar:2019aa, Gibson:20, Erkmen:10, Shapiro:2008, Shapiro:2012aa, Pelliccia:2016, Radwell:14, Bian:2016aa, Pittman:1995, Shih:24, Chan:2008}.
In these frameworks, the target is illuminated by a series of spatially structured patterns, and the corresponding spatially integrated intensities are recorded with a single photodetector. The original image is then computationally reconstructed, with modern implementations often leveraging compressive sensing \cite{Shapiro:2008, Candes:2008, Duarte:2008, Katz:2009, Gong:2012, Zhao:2012, Han:2018, Donoho:2006, Tian:2022} or deep-learning approaches \cite{Lyu:2017aa, He:2018aa, Wang:19, Wu:20, Yu:22} to improve reconstruction fidelity using fewer measurements.

Despite these advantages, SPI/GI has traditionally been limited in temporal resolution, as the speed of illumination-pattern switching remains a critical bottleneck.
To address this limitation, various high-speed SPI or GI strategies have been explored \cite{Hahamovich:2021aa,Lin:25,Kilcullen:2022aa,Xu:18,Huang:2022,Johnstone:24,Zhao:2019,Stojek:2022,Kanno:20,Jiang:22}.  
For example, cyclic Hadamard mask projection using a rotating disk has achieved megahertz (MHz)-rate operation \cite{Hahamovich:2021aa, Lin:25}. More advanced techniques have increased projection rates to 14.1 MHz by integrating laser-scanning hardware with a digital micromirror device (DMD) \cite{Kilcullen:2022aa}. LED- and microLED-based configurations have demonstrated purely electronic modulation in the MHz range, enabling video-rate imaging at kiloframes per second (kfps) \cite{Xu:18, Huang:2022, Johnstone:24, Zhao:2019, Stojek:2022}. 
However, all these approaches exhibit inherent limitations-- whether relying on mechanical motion, restricted modulation bandwidth, or applicability confined to periodic dynamics--and none have operated beyond frame rates of a few tens of megahertz, rendering them insufficient for capturing microsecond-scale dynamic phenomena.

An optical illumination switching approach without any mechanical switching is promising to address the limitation issue and achieve faster image acquisition. 
In our previous work, we proposed an ultrafast random speckle illumination approach at Gigahertz switching rates \cite{Hanawa:22, Yamaguchi:2023aa} and demonstrated that this switching approach has allowed for high-speed image recognition and GI at 20 megaframes per second (Mfps). 
Very recently, Wan {\it et al.} have demonstrated 20 Mfps single-pixel imaging using a scan-less technique based on random speckles assisted by dual optical frequency combs \cite{Wan:2025}. While these speckle-based illumination approaches have potential for faster image acquisition, these previous studies rely on supervised learning for image reconstruction using training data samples, which makes it challenging to capture previously unseen events. 

Here, we propose an ultrafast GI approach based on 25 GHz random speckle illumination combined with a training-data-free reconstruction algorithm, enabling accurate reconstruction of previously unseen dynamic events.
A remarkable feature of the proposed approach is its inherent ability to incorporate wavelength-division multiplexing (WDM) to parallelize the projection of multiple independent patterns across distinct wavelength channels, 
achieving an effective switching rate of $25 \times 5 = 125$ GHz with five-wavelength lasers. 
A high-quality image can be reconstructed using a self-supervised deep neural network without relying on training samples. We experimentally demonstrate that the proposed approach is capable of reconstructing $28 \times 28$-pixel images at $100$ Mfps, corresponding to a spatial-temporal information flux (STIF) of $28 \times 28 \times 100 = 78.4$ gigapixels per second (Gpps) across five wavelengths. This rate is much higher than that reported in previous work \cite{Wan:2025}. The system performance reveals its potential for observing microsecond-scale dynamics in physics, chemistry, and biology using only a few photodetectors. By bridging the gap between extreme temporal resolution and computational imaging flexibility, this approach defines a new paradigm for ultrafast single-pixel imaging (SPI) and GI.

% \section{SMSPI}\label{sec2}
\section{Concept and Methods}\label{sec2}

% \subsection{SMSPI system}
\subsection{System overview}
\label{sec2.1}
\begin{figure}[htbp]
\centering\includegraphics[bb=0 0 576 313,width=12cm]{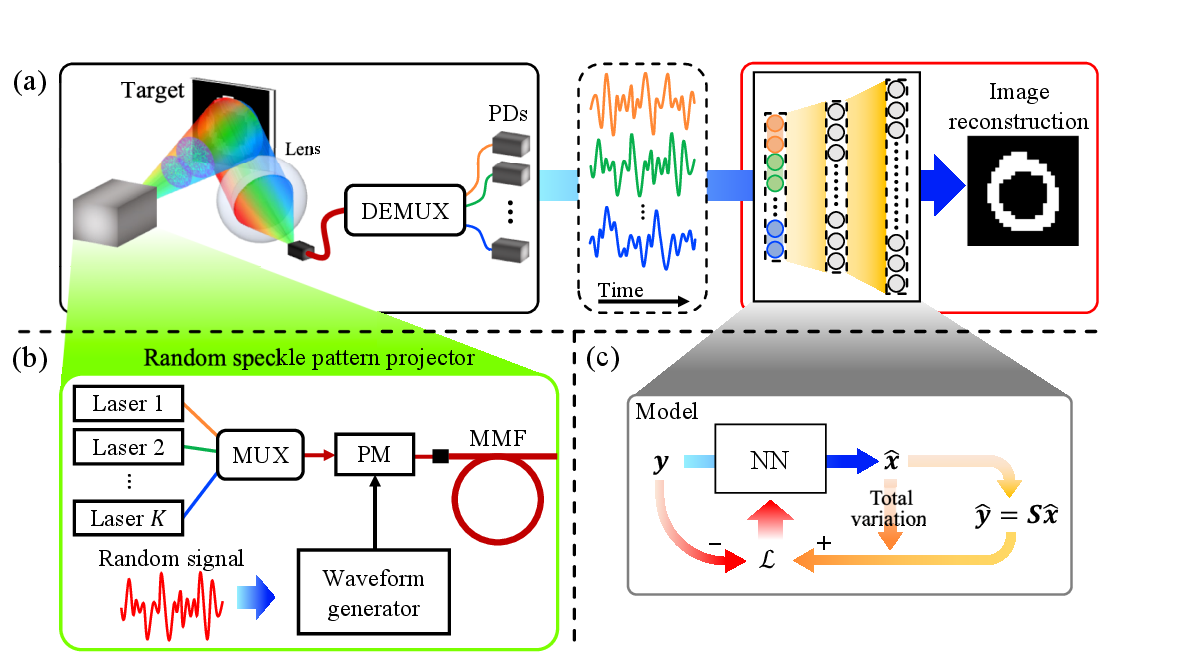}
\caption{Conceptual schematic of the proposed wavelength-multiplexed ghost imaging (WDM-GI) system. 
(a) Schematic of the WDM-GI system. 
DEMUX: demultiplexer; PDs: photodetectors. 
(b) Random speckle pattern projector. MUX: multiplexer; PM: phase modulator; MMF: multimode fiber. 
(c) Reconstruction model. NN: neural network.}
\label{fig2-1}
\end{figure}

Figure \ref{fig2-1} (a) shows the conceptual schematic of the proposed wavelength-division-multipled ghost imaging (WDM-GI) system, which mainly consists of two units: random speckle pattern projector and receiver units.

The schematic of the projector unit is shown in Fig.\ref{fig2-1}~(b). This unit comprises multi-wavelength lasers, a multiplexer (MUX, arrayed waveguide grating in our experiment), a phase modulator (PM), and a multimode fiber (MMF). Laser light from the multi-wavelength sources is combined using the MUX and simultaneously phase-modulated using the PM with a pseudorandom sequence generated by a waveform generator (WG) operating at 25 Gigasamples per second (GSa/s). The modulated light is introduced into an MMF with a length of 20~m and a core diameter of 200~$\mu$m, generating temporally varying speckle patterns for each input wavelength (see Sec.~1 in Supplement 1). 
Speckle mask patterns corresponding to different wavelengths are simultaneously projected onto the target object. 

In the receiver unit, reflected light from the target is collected by a lens, demultiplexed by a demultiplexer (DEMUX), and detected by multiple photodetectors (PDs), which convert the signals into the time domain for each wavelength [Fig.~\ref{fig2-1}(a)]. The measured time-domain signals are then used for image reconstruction. Details of the experimental setup are provided in Sec.~2 of Supplement 1.

The aforementioned image-to-temporal conversion process, combined with WDM, is formulated as follows. Let the target image and speckle mask pattern at wavelength $\lambda_k$ ($k = 1,2,\cdots,K$) as $\boldsymbol{x} = (x_1, \ldots, x_j, \ldots, x_N)^\top \in \mathbb{R}^N$
and $\boldsymbol{s}^{\lambda_k}(t_i) = (s^{\lambda_k}_1(t_i),\cdots,s^{\lambda_k}_N(t_i))^{\top}$, respectively, where $N$ is the total number of pixels, and discrete time is given as $t_i = i \Delta t$ $(i=1, 2,..., M)$. 
$\Delta t$ represents the time resolution, and $M$ determines the exposure time used for the image reconstruction: $T= M \Delta t$. $K$ is the number of laser wavelengths.  
For the formulation, we define an $M\times N$ matrix $\boldsymbol{S}^{\lambda_k} = ( \boldsymbol{s}^{\lambda_k} (t_1), \cdots, \boldsymbol{s}^{\lambda_k}(T) )^{\top}$ at wavelength $\lambda_k$. 
The measured time-domain signal at time $t_i$ is expressed as a vector $\boldsymbol{y}^{\lambda_k} = (y^{\lambda_k}(t_1), \dots, y^{\lambda_k}(t_i), \dots, y^{\lambda_k}(T))^\top \in \mathbb{R}^M$.
The total measurement matrix $\boldsymbol{S}$ and the measurement vector $\boldsymbol{y}$ are defined as follows:
\begin{align}
\boldsymbol{S} =
\begin{pmatrix}
 \boldsymbol{S}^{\lambda_1} \\
 \boldsymbol{S}^{\lambda_2} \\
 \vdots \\
 \boldsymbol{S}^{\lambda_{K-1}} \\
 \boldsymbol{S}^{\lambda_K},
\end{pmatrix}
\quad
\boldsymbol{y} =
\begin{pmatrix}
 \boldsymbol{y}^{\lambda_1} \\
 \boldsymbol{y}^{\lambda_2} \\
 \vdots \\
 \boldsymbol{y}^{\lambda_{K-1}} \\
 \boldsymbol{y}^{\lambda_K},
\end{pmatrix}
%\in \mathbb{R}^{KM}.
\label{eq:SandY}
\end{align}
where the sizes of $\boldsymbol{S}$ and $\boldsymbol{y}$ are $KM \times N$ and $KM$, respectively.
Since $\boldsymbol{y}$ can be represented by the measurement matrix $\boldsymbol{S}$ and the target image $\boldsymbol{x}$ as follows, 
\begin{align}
 \boldsymbol{y} = \boldsymbol{S}\boldsymbol{x}, \label{eq_ysx}
\end{align}
the target image $\boldsymbol{x}$ can be reconstructed as $\hat{\boldsymbol{x}} = \boldsymbol{S}^{\dag}\boldsymbol{y}$, where $\boldsymbol{S}^{\dag}$ represents the pseudo-inverse matrix of $\boldsymbol{S}$ for $KM \gg N$. However, the reconstruction approach based on the pseudo-inverse matrix may lead to a low-quality image when the matrix $\boldsymbol{S}$ is computationally low rank. To address this ill-posed issue, we instead employ a neural network approach, as shown in the next subsection.  

\subsection{Reconstruction Algorithm}
\label{sec2.2}
The image vector $\boldsymbol{x}$ is reconstructed using a self-supervised learning technique based on Eq.~(\ref{eq_ysx}), as shown in Fig.\ref{fig2-1} (c). While numerous reconstruction algorithms have been effective, we adopt Ghost Imaging using Deep neural-network Constraint (GIDC) \cite{Wang:2022aa} because it provides excellent reconstruction performance even for correlated speckle mask patterns \cite{Wang:2022aa}. Additionally, we found that the GIDC is well-suited for a random mask matrix with an exponential distribution, such as the speckle mask matrix $\boldsymbol{S}$, compared to other reconstruction algorithms such as Unrolling Convolutional Neural Network (CNN) \cite{Quan:2022,Diamond2017Unrolled}.
Furthermore, the GIDC yields better reconstruction results, particularly for larger image data. See Sections 3 and 4 in Supplement 1 for the detailed discussion.

Figure~\ref{fig2-2} depicts the overall architecture of the GIDC. Firstly, a reconstruction image is created based on Eq.~(\ref{eq_ysx}) as: $\hat{\boldsymbol{x}} = \boldsymbol{S}^{\dag} \boldsymbol{y}$.
Secondly, the image is fed to a CNN, which refines the reconstruction by leveraging its denoising capability and converts it into image data $\hat{\boldsymbol{x}}$. 
Then, $\hat{\boldsymbol{y}} = \boldsymbol{S}\hat{\boldsymbol{x}}$ is computed. The GIDC is trained to minimize the following loss function:
\begin{align}
\mathcal{L} = \left|
\boldsymbol{\hat{y}} - \boldsymbol{y} 
\right|^2 + R(\hat{\boldsymbol{x}}),
\label{eq:lossfunction}
\end{align}
where $R(\hat{\boldsymbol{x}})$ denotes the total variation (TV), which contributes to smoothing the output images \cite{Liu:2019, Rudin:1992}.

\begin{figure}[htbp]
\centering\includegraphics[bb=0 0 523 306,width=10cm]{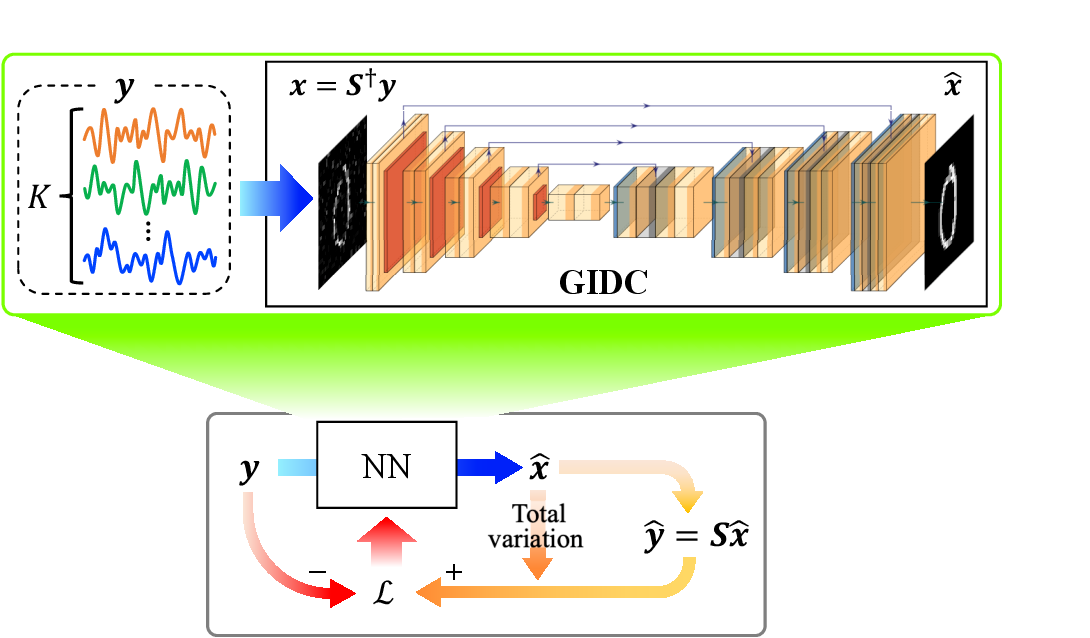}
\caption{Reconstruction model.
The network first computes a simplified reconstructed image from the pseudo-inverse of the measurement matrix $\boldsymbol{S}$, then refines the image using a CNN to suppress noise and enhance structural fidelity. The U-Net architecture is trained such that the output $\hat{\boldsymbol{y}} = \boldsymbol{S}\hat{\boldsymbol{x}}$ matches the target $\boldsymbol{y}$.}
\label{fig2-2}
\end{figure}

\subsection{Spatiotemporal measurement of speckle dynamics}\label{sec2.3}
To reconstruct the image based on the measurement matrix $\boldsymbol{S}$, the spatiotemporal variation of random speckle patterns generated at 25~GSa/s must be measured; however, GHz-rate spatiotemporal measurement is difficult using a conventional image sensor. To address this issue and estimate the matrix $\boldsymbol{S}$, multiple images $\boldsymbol{x}_r$ $(r = 1,2, \cdots, N_R)$, designed as random patterns, were displayed on a DMD, and the speckle pattern $\boldsymbol{s}^{\lambda_k}(t)$ at wavelength $\lambda_k$ was projected onto the image. Then, the corresponding time-domain signals, $y_r^{\lambda_k}(t_i) = (\boldsymbol{s}^{\lambda_k}(t_i))^\top \boldsymbol{x}_r$, were acquired. 
For improved reconstruction, the inverted counterparts of the target images, $\bar{\boldsymbol{x}}_r$, were also displayed, and the corresponding signals $\bar{y}_r^{\lambda_k}(t_i)$ were acquired.
Here let $\boldsymbol{X}_{\rm cal}$ and $\boldsymbol{Y}_{\rm cal}^{\lambda_k}$ be $(\tilde{\boldsymbol{x}}_1,\cdots,\tilde{\boldsymbol{x}}_{N_R})$ and 
$(\tilde{\boldsymbol{y}}^{\lambda_k}_1,\cdots,\tilde{\boldsymbol{y}}_{N_R}^{\lambda_k})$, respectively, where 
$\tilde{\boldsymbol{x}}_r = \boldsymbol{x}_r-\bar{\boldsymbol{x}}_r$ and 
$\tilde{\boldsymbol{y}}^{\lambda_k}_r =\left(y_r^{\lambda_k}(t_1)-\bar{y}_r^{\lambda_k}(t_1), \cdots, y_r^{\lambda_k}(T)-\bar{y}_r^{\lambda_k}(T)\right)^\top$. 
The measurement matrix $\boldsymbol{S}^{\lambda_k}$ at wavelength $\lambda_k$ is inferred as follows:
\begin{align}
 \boldsymbol{S}^{\lambda_k} =  \boldsymbol{Y}_{\rm cal}^{\lambda_k}  \boldsymbol{X}_{\rm cal}^{\dag}. 
\end{align}
Consequently, the total measurement matrix $\boldsymbol{S}$ is represented as $(\boldsymbol{S}^{\lambda_1}, \cdots,{\boldsymbol{S}}^{\lambda_K})$. 

\begin{figure}[htbp]
\centering\includegraphics[bb=0 0 581 425,width=12cm]{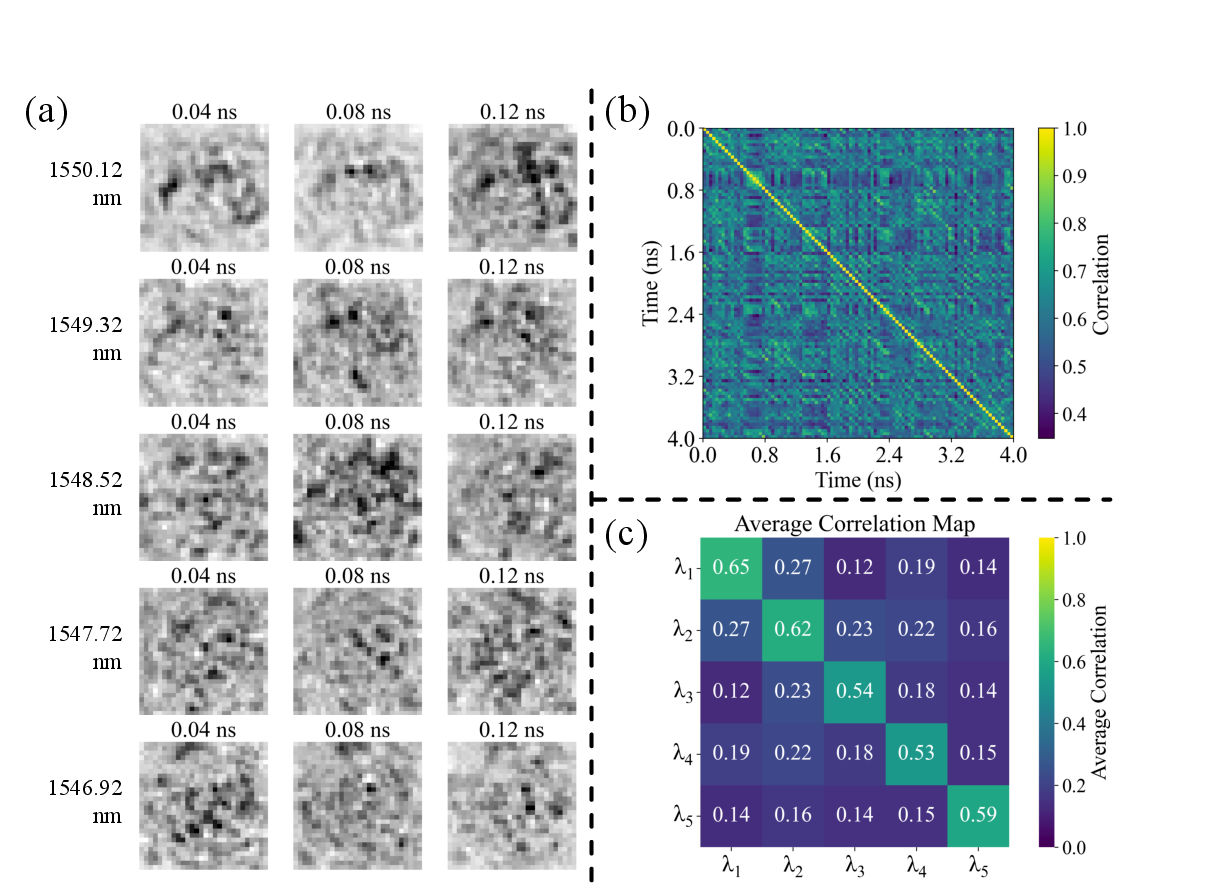}
\caption{
(a) Measured random speckle patterns for $\lambda_1~=~$1550.12, $\lambda_2~=~$1549.32, $\lambda_3~=~$1548.52, $\lambda_4~=~$1547.72, and $\lambda_5~=~$1546.92 nm.
(b) Correlation matrix between the measured speckle patterns for a single wavelength (1546.92 nm). Both axes are expressed in time (ns).
(c) Average correlation matrix across five wavelengths. 
}
\label{fig2-3}
\end{figure}

Figure~\ref{fig2-3}(a) illustrates the speckle patterns measured using the aforementioned method for each wavelength at intervals of $\Delta t = 0.04$~ns. 
The reconstructed pattern size was set to $N = 28 \times 28$ pixels. 
In the measurement, we used 2,500 random patterns ($N_R = 2{,}500$). 
The number of patterns, $N_R$, was chosen to be larger than the number of pixels, $N$, to reduce measurement error. See Sec.~5 in Supplement 1 for details.
Figure~\ref{fig2-3}(b) shows the cross-correlation matrix $C_{\lambda_k}(j\Delta t)~=~\langle (\boldsymbol{s}^{\lambda_k}(t))^\top\boldsymbol{s}^{\lambda_k}(t+j\Delta t)\rangle_T$ between the speckle patterns for wavelength $\lambda_k = 1550.12~$nm, where $\langle\cdot\rangle_T$ denotes the time average. The correlation drops sharply for $\Delta t~=~$0.04 ns. The speckle correlation for different wavelengths was evaluated using the average correlation matrix $C_{\lambda_k,\lambda_{k'}}~=~(1/T)\sum_jC_{\lambda_k}(j\Delta t)C_{\lambda_{k'}}(j\Delta t)$ [Fig.~\ref{fig2-3} (c)]. The diagonal elements represent intra-wavelength correlations, while the off-diagonal elements correspond to inter-wavelength correlations and exhibit very low values, confirming that speckle patterns at different wavelengths are mutually independent.

% \subsection{Wavelength-Multiplexed High-Speed Mask Pattern Projector}

\section{Results}\label{sec3}

\subsection{Image reconstruction}
\label{sec3.1}
\begin{figure}[htbp]
\centering\includegraphics[bb=0 0 425 573, width=11cm]{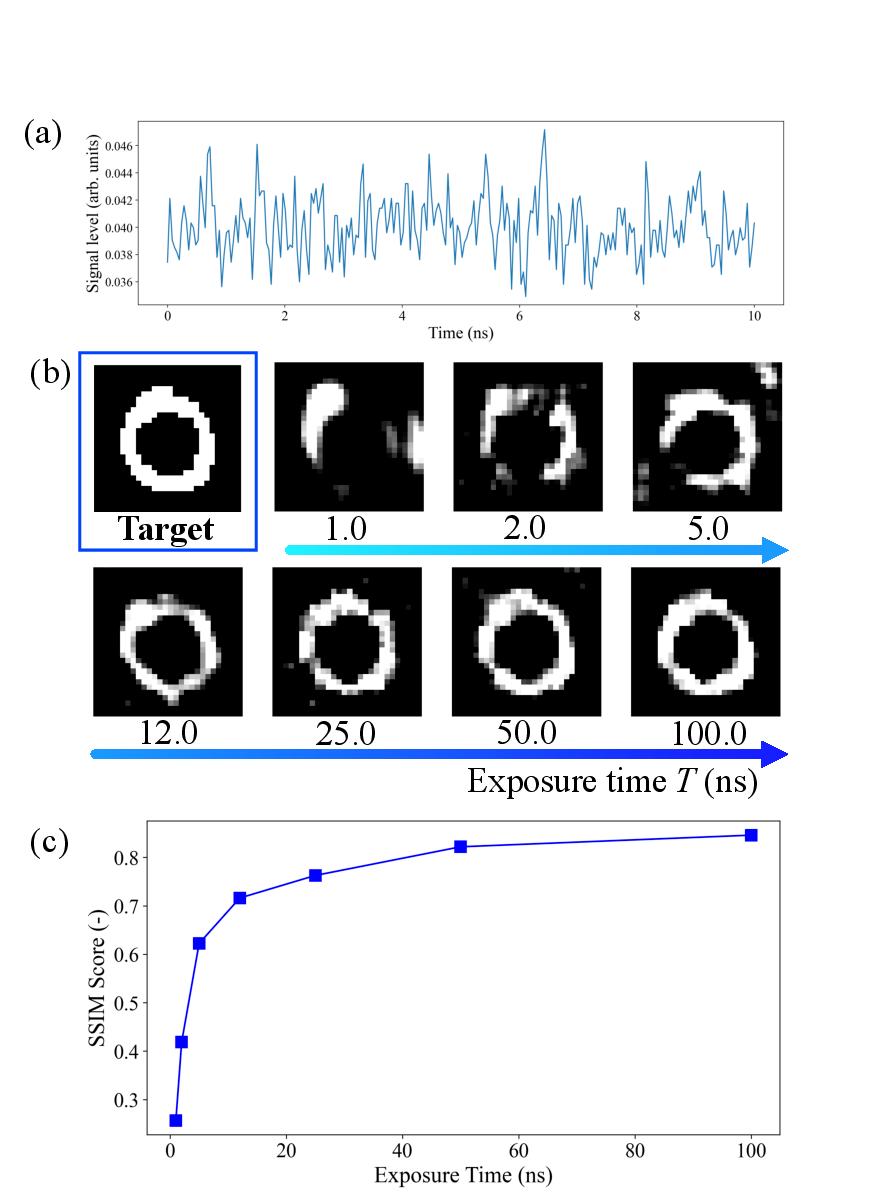}
\caption{Reconstruction results with varying exposure times ($K$~=~1). 
(a) Measured time-domain signal and (b) its reconstructed images. 
Exposure time $T$: 1.0 - 100.0~ns. $\lambda = 1546.92$ nm. 
(c) SSIM as a function of $T$.}
\label{fig3-1}
\end{figure}

To verify the effectiveness of the proposed GI approach, we used the MNIST handwritten digit images with $28 \times 28$ pixels as target images and displayed the images on the DMD.  
Figures~\ref{fig3-1}(a) and \ref{fig3-1}(b) show the measured time-domain signal for the digit ``0'' image and its reconstruction results for exposure times $T = 1$ -- $100$~ns in the single-wavelength case ($K$~=~1). 
The visibility of the reconstruction improves with increasing exposure time $T$, whereas noise and edge loss are pronounced under short-exposure conditions.
The structural similarity index measure (SSIM) in Fig.~\ref{fig3-1}(c) shows a monotonic increase with $T$, indicating that the performance limitation of the single-wavelength configuration becomes apparent under the short-exposure regime ($T <10$~ns).

\begin{figure}[htbp]
\centering\includegraphics[bb=0 0 468 536, width=11cm]{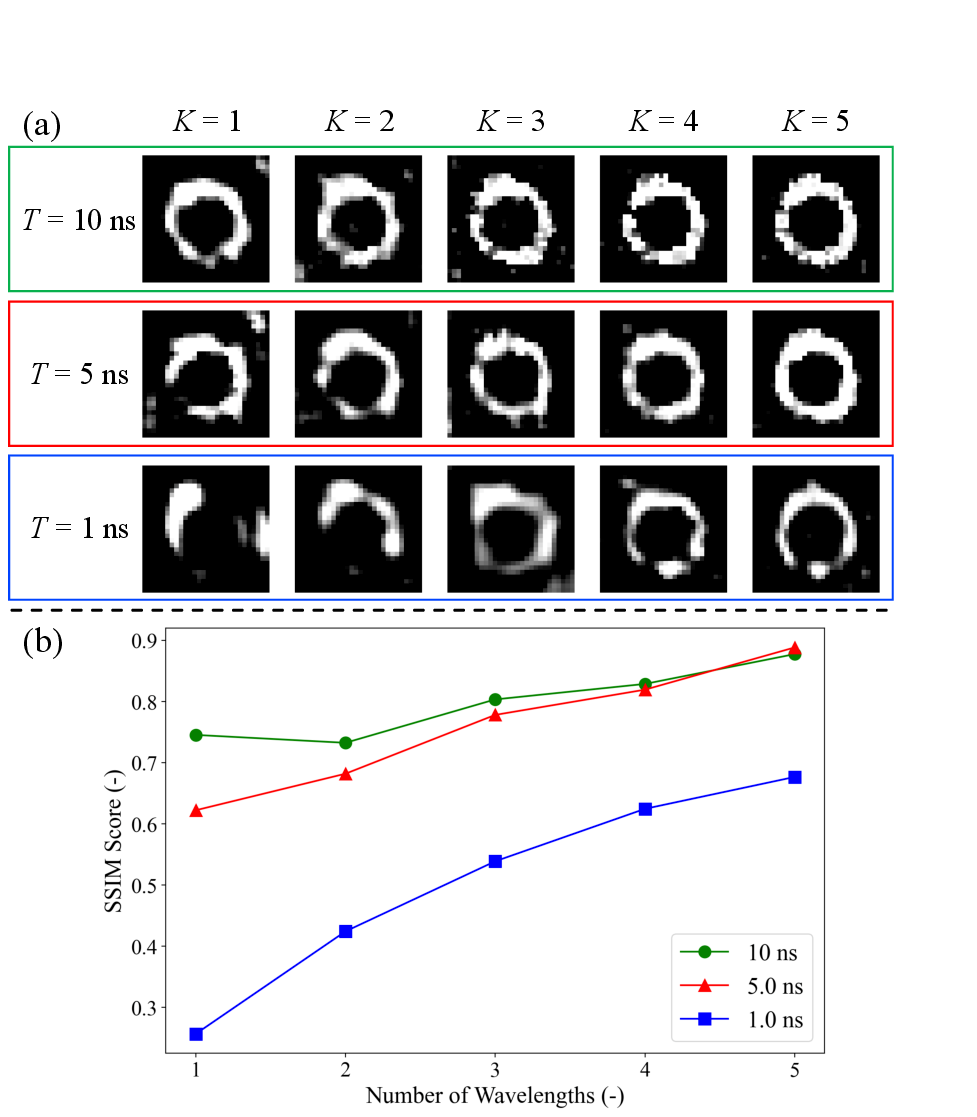}
\caption{Reconstruction results based on the WDM-GI. 
Exposure time $T$: 1.0, 5.0, 10.0~ns, number of wavelengths $K$ = 1 - 5, (a) Reconstructed images, (b) SSIM versus $K$ for each $T$. 
The gain from increasing $K$ is larger at shorter $T$. 
Our WDM-GI enables high-quality reconfiguration with a short exposure time $T \approx 10$ ns, corresponding to 100 Mfps.}
\label{fig3-2}
\end{figure}

\begin{figure}[htbp]
\centering\includegraphics[bb=0 0 724 197, width=13cm]{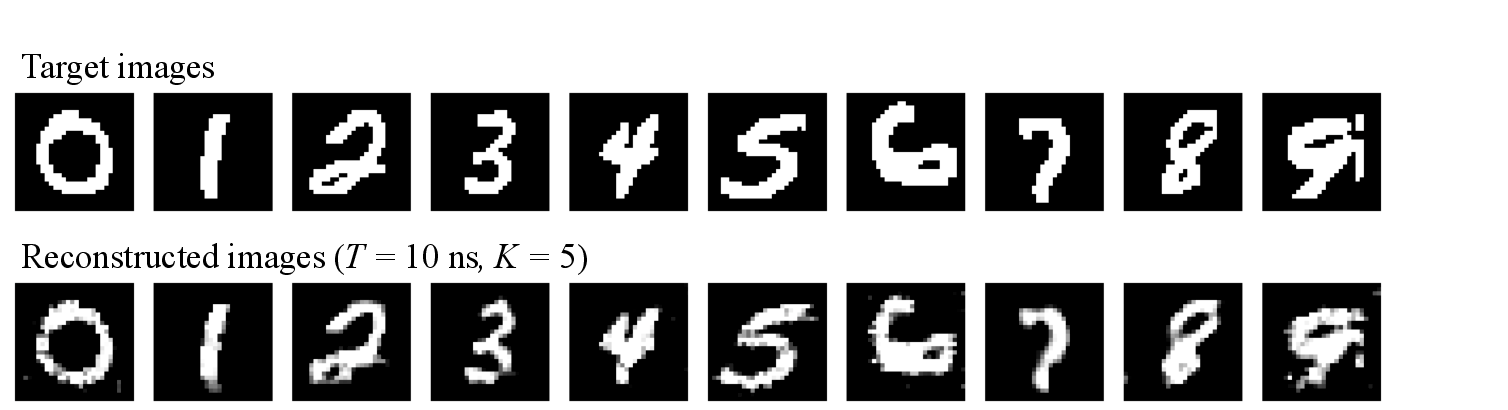}
\caption{Ground truth and reconstructed images. $T~=~10$ ns and $K~=~5$.
The SSIM values for digits ``0'' through ``9'' are 0.8773, 0.9429, 0.8507, 0.8568, 0.8627, 0.8328, 0.7814, 0.8607, 0.8491, and 0.7467, respectively.}
\label{fig3-3}
\end{figure}

The WDM-GI can be used for improving the reconstruction quality under short-exposure conditions. To verify the effectiveness, we set $T =1.0, 5.0$, and $10.0$~ns and varied the number of wavelengths $K = 1, \cdots, 5$. 
Figures~\ref{fig3-2}(a) and \ref{fig3-2}(b) show the reconstruction results and the SSIM, respectively. 
Across all exposure times $T$, the SSIM improves monotonically with $K$, with substantial improvement at the ultra-short exposure time $T=1.0$~ns [Fig.~\ref{fig3-2}(b)]. Simultaneous projection of uncorrelated speckle patterns increases pattern diversity and improves the reconstruction quality. For $K \ge 4$, the SSIM exceeds 0.8 with the exposure time $T \ge 10$ ns, suggesting the possibility of 100 Mfps imaging.
Examples of other reconstructed images for $T~=~10$ ns and $K~=~5$ are shown in Fig.~\ref{fig3-3}. The mean SSIM was approximately 0.85.

\subsection{Video recording of a microsecond-scale switching event.}
\label{sec3.2}
\begin{figure}[htbp]
\centering\includegraphics[bb=0 0 386 409,width=11cm]{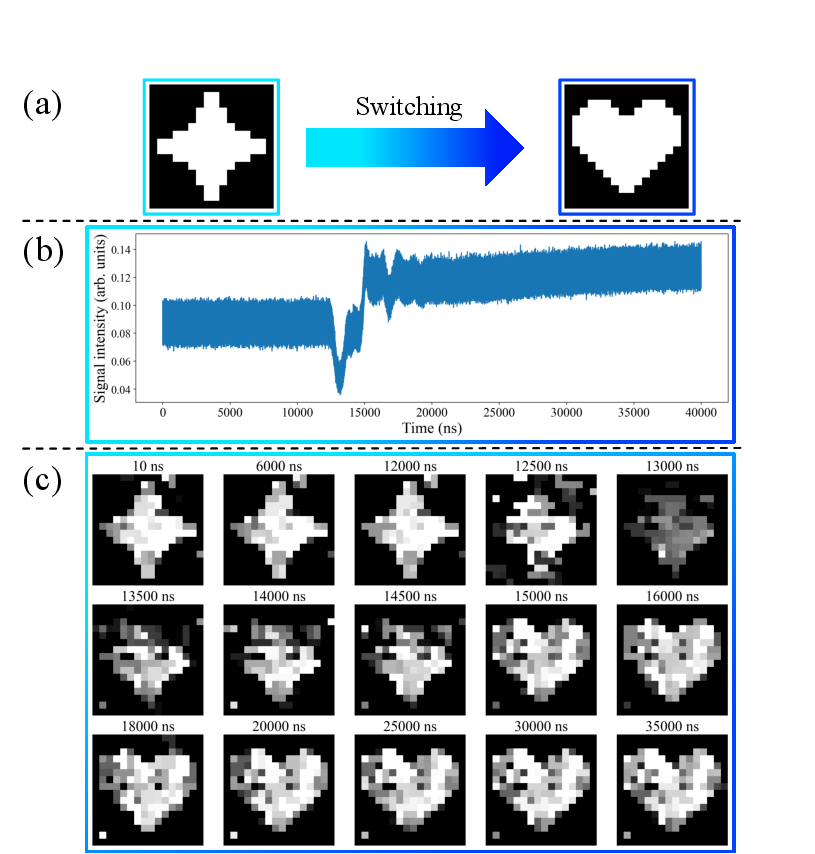}
\caption{Video recording of a microsecond-scale switching event (diamond to heart) at 100 Mfps. (a) Target images. (b) Acquired time-domain signal. 
(c) Representative reconstructed images ($K = 5$). The images were reconstructed from the acquired time-domain signals with a 10-ns interval. To visualize the transition from the diamond-shaped to the heart-shaped pattern, only selected reconstructed images are shown here.
}
\label{fig3-4}
\end{figure}

Taking advantage of the proposed WDM-GI, the temporal evolution of a microsecond-scale dynamic event can be visualized. For the proof-of-concept demonstration, we chose a high-speed image-switching event from a diamond-shaped image with 16$\times$16 pixels on the DMD into a heart-shaped image [Fig.~\ref{fig3-4}(a)].
To reconstruct the images appearing during the switching event, the acquired continuous time-domain signals were divided into equal time segments of 10~ns. The exposure time $T$ was set to $10$~ns, and $K~=~5$ wavelengths were employed. %The laser wavelengths used in this demonstration were the same as those mentioned in Subsec.~\ref{sec2.2}.

Figure~\ref{fig3-4}(b) shows the time-domain signal measured during the switching event under speckle mask projection at a wavelength of $1550.12$~nm. The time-domain signal changed significantly during the target image transition ($t = 1.2$--$1.6 \mu$s). 
Figure~\ref{fig3-4}(c) shows some typical examples of the reconstructed image sequence recorded at $10$~ns intervals using our WDM-GI approach. 
The reconstructed images show the smooth transition from the diamond to the heart shapes. Small fluctuations after the transition ($t= 1.5$--$2.0$) can be attributed to a relaxation oscillation after the mechanical mirror switching on the DMD.
With a reconstruction interval of $T=10$~ns, the effective frame rate was $100$~Mfps.
This result demonstrates that our ultrafast GI technique can capture microsecond-scale dynamics as a continuous frame sequence (i.e., a video).

\section{Discussion}\label{sec4}
\begin{table*}[t]
\small
\centering
\caption{
Comparison table.
$^\dag$ Effective STIF values for $K~=~5$. (78.4/5 = 15.68 Gpps and 784/5 = 156.8 Gpps for $K~=~1$.) $^{\dag\dag}$ A physics-driven fine-tuning to refine a pretrained model. $^\ddag$ 2 Mfps for a periodic image.
}

%\begin{tabularx}{\textwidth}{|c|c|c|c|c|c|c|}
\begin{tabular}{|c|c|c|c|c|c|c|}
\hline
Ref. & Mask switching & Requires & Switching rate & Image size & Frame rate & STIF  \\
     & method         & training data & (GHz)          & (pixels)   & (Mfps)     & (Gpps)\\
\hline
\textbf{This} & Optical & No &  25 & $28 \times 28$ & \textbf{100} & 78.4$^\dag$ \\%78.4 \\
\textbf{work} &  & Yes & &  & \textbf{1,000} & 784$^\dag$\\%784 \\
\hline 
\cite{Hahamovich:2021aa} & Mechanical & No & 2.4$\times 10^{-3}$ & $100 \times 100$ & $7.2\times 10^{-5}$ & 0.00072\\
\hline
\cite{Lin:25} & Mechanical & -$^{\dag\dag}$ & 1$\times 10^{-3}$ & $71 \times 73$ & $1.926\times 10^{-3}$ & 0.00998\\
\hline
\cite{Kilcullen:2022aa} & Mechanical & No & 1.41$\times 10^{-2}$ & $101 \times 103$ & $1.2\times 10^{-2}$ & 0.125\\
\hline
\cite{Xu:18} & Electronic & No & 1$\times 10^{-3}$ & $32 \times 32$ & $1.0\times 10^{-3}$ & 0.00102\\
\hline
\cite{Huang:2022} & Electronic & No & 1.25$\times 10^{-2}$ & $32 \times 32$ & $2.5\times 10^{-2}$ & 0.0256 \\
\hline
\cite{Johnstone:24} & Electronic & No & 4$\times 10^{-4}$ & $128 \times 128$ & $8.0\times 10^{-4}$ & 0.013\\
\hline
\cite{Zhao:2019} & Electronic & No & 0.1 & $8 \times 8$ & 1.4 & 0.0896\\
\hline
\cite{Stojek:2022} & Electronic & No & 2.27$\times 10^{-5}$ & $1024 \times 768$ & $7.0\times 10^{-6}$ & 0.0055\\
\hline
\cite{Kanno:20} & Optical & No & -- & -- & $3.2\times 10^{-2}$ & --\\
\hline
\cite{Jiang:22} & Optical & No & -- & -- & 2.0$^\ddag$ & --\\
\hline
\cite{Yamaguchi:2023aa} & Optical & Yes & 25 & $28 \times 28$ & 50 & 39.2\\
\hline
%\cite{Wan:2025} & Optical & \textbf{S} & 18 & $28 \times 28$ & 20 & 15.68\\
\cite{Wan:2025} & Optical & Yes & Scan-less & $28 \times 28$ & 20 & 15.68\\
\hline
%\end{tabularx}
\end{tabular}
\label{table}
\end{table*}

To clarify the features of our approach, we compare it with existing high-speed SPI/GI methods. Table~\ref{table} summarizes the mask-switching methods, the requirement for training datasets in image reconstruction, switching rate, image size, and frame rate. For fair comparison, we also present the estimated spatial-temporal information flux (STIF) \cite{Wan:2025}, defined as the product of the number of image pixels and the frame rate (expressed in gigapixels per second). 

The mask generation techniques used in previous studies can be broadly classified into mechanical, electronic, and optical types. Mechanical systems typically rely on rotating disks or scanning apertures, where inertia and motor speed restrict the mask-switching rate to a few kilohertz to tens of megahertz. 
Consequently, these approaches achieve frame rates no higher than $\sim10^{-2}$~Mfps and STIF values on the order of $10^{-3}$--$10^{-1}$~Gpps~\cite{Hahamovich:2021aa, Lin:25, Kilcullen:2022aa}. 
Electronic switching approaches (using DMDs, LEDs, or micro-LED arrays) are constrained by the electronic rise time of the modulators and the need for sequential pattern projection. 
As a result, typical frame rates range from $10^{-3}$ to 1~Mfps, and the STIF remains below a few~$\times 10^{-2}$~Gpps, except when very small image formats are used~\cite{Xu:18, Huang:2022, Johnstone:24, Zhao:2019, Stojek:2022}. Unlike mechanical or electronic approaches, the optical switching approach is not limited by moving parts or electronic rise times, enabling much faster switching rates and frame rates~\cite{Kanno:20, Jiang:22, Yamaguchi:2023aa, Wan:2025}. In particular, speckle-switching approaches can achieve tens of Gpps in the STIF \cite{Yamaguchi:2023aa, Wan:2025}. However, in these previous studies, image reconstruction relies on supervised learning using training data samples, making it difficult to reconstruct unseen images.

In contrast, our approach utilizes a training-data-free algorithm and reconstructs target images directly from the time-domain signal using a 25~GHz random-speckle pattern projection. The STIF can achieve 78.4~Gpps for $K~=~5$, which is higher than previous approaches. Consequently, our WDM-GI is capable of 100~Mfps imaging of previously unseen dynamic events.

When prior knowledge of a target image is available, a supervised-learning approach with labeled data can be applied. In this case, our WDM-GI system scales up to 1000~Mfps and an STIF of 784~Gpps, illustrating that additional training data can unlock an order-of-magnitude improvement in throughput. 
See Sec.~6 in Supplement 1 for further details. 

Taken together, Table~\ref{table} highlights that our method outperforms previous approaches by a wide margin, achieving a performance regime not previously reported. 
This establishes a new operational domain for SPI/GI. 
Looking ahead, increasing the number of wavelength channels and refining speckle diversity could further enhance throughput without sacrificing temporal resolution, suggesting a clear path toward scalable ultrafast computational imaging.

Despite these advantages, several directions remain for improving reconstruction performance. 
A major bottleneck lies in quantization error during time-domain signal acquisition. 
The current oscilloscope provides only eight-bit resolution, which degrades reconstructed image quality (see Sec.~7 in Supplement 1). 
Employing higher-bit-depth acquisition and incorporating quantization-aware training could further suppress this error and lead to improved reconstruction quality. 

To enable the reconstruction of higher-resolution images, the speckle size, which determines image resolution, should be reduced, for example, by introducing an additional diffuser or a spatial light modulator for programmable optimization. 
Furthermore, less-correlated speckle patterns can be generated by improving coupling into the MMF so that more guided modes are excited by the input signal (Sec.~1 in Supplement 1).
The stability of speckle patterns is also a crucial factor for reliable reconstruction. 
Long-term temporal drift or temperature-induced variations in the optical path can degrade reconstruction quality. 
Implementing active stabilization techniques, such as feedback control of optical alignment or environmental compensation, would improve the quality of measured patterns. 

Finally, the demultiplexing process at the receiver plays a key role in separating signals corresponding to different wavelength channels. 
By incorporating wavelength-selective components or scattering units, it becomes possible to achieve a more efficient GI system capable of operating across multiple wavelength channels (Sec.~8 in Supplement 1).

\section{Conclusion}\label{sec5}
In this study, we proposed an ultrafast GI approach combined with a self-supervised, training-data-free decoding model to capture previously unseen images.
The proposed optical switching technique enables the parallel generation of multiple independent speckle mask patterns.
The resulting STIF exceeded 78.4 Gpps for $K~=~5$, which is much higher than those in previous approaches and enabled video recording with $100$~Mfps. 
These results indicate that our approach overcomes the limitations of conventional approaches and establishes an effective framework for ultrafast, high-throughput GI.

This method offers a unique form of GI/SPI with advantages in terms of temporal resolution, scalability, and ease of implementation.
Future work includes increasing the multiplicity through the expansion of wavelength channels, optimizing the stability and diversity of speckle patterns, and applying the method to dynamic observation of three-dimensional structures, among other potential developments.
The demonstrated performance reveals its potential for imaging sub-microsecond-scale dynamics in physics, chemistry, and biology. By bridging the gap between extreme temporal resolution and computational imaging flexibility, this approach defines a new paradigm for ultrafast SPI and GI.

%\section*{Acknowledgements}
\begin{acknowledgements}
This work was supported in part by 
JSPS KAKENHI (Grant Nos.~JP22H05198, JP23K28157),
JST CREST (Grant No.~JPMJCR24R2), and 
JKA promotion funds from KEIRIN RACE (No. 2024M-515).
\end{acknowledgements}


%\bibliography{OPTICA_ref}
%\bibliographystyle{apsrev}
%\bibliography{apsrev.bst}
%\bibliography{/Users/sunada/Dropbox/My_Refs_db}
%\bibliography{ms_dfa-adjoint_ver9.bbl}

\begin{thebibliography}{10}
\newcommand{\enquote}[1]{``#1''}

\bibitem{Edgar:2019aa}
M.~P. Edgar, G.~M. Gibson, and M.~J. Padgett, \enquote{Principles and prospects
  for single-pixel imaging,} {{Nature Photonics}}
  \textbf{13}, 13--20 (2019).

\bibitem{Gibson:20}
G.~M. Gibson, S.~D. Johnson, and M.~J. Padgett, \enquote{Single-pixel imaging
  12 years on: a review,} {{Opt. Express}} \textbf{28},
  28190--28208 (2020).

\bibitem{Erkmen:10}
B.~I. Erkmen and J.~H. Shapiro, \enquote{Ghost imaging: from quantum to
  classical to computational,} {{Adv. Opt. Photon.}}
  \textbf{2}, 405--450 (2010).

\bibitem{Shapiro:2008}
J.~H. Shapiro, \enquote{Computational ghost imaging,}
  {{Phys. Rev. A}} \textbf{78}, 061802 (2008).

\bibitem{Shapiro:2012aa}
J.~H. Shapiro and R.~W. Boyd, \enquote{The physics of ghost imaging,}
  {{Quantum Information Processing}} \textbf{11}, 949--993
  (2012).

\bibitem{Pelliccia:2016}
D.~Pelliccia, A.~Rack, M.~Scheel, \emph{et~al.}, \enquote{Experimental x-ray
  ghost imaging,} {{Phys. Rev. Lett.}} \textbf{117},
  113902 (2016).

\bibitem{Radwell:14}
N.~Radwell, K.~J. Mitchell, G.~M. Gibson, \emph{et~al.}, \enquote{Single-pixel
  infrared and visible microscope,} {{Optica}} \textbf{1},
  285--289 (2014).

\bibitem{Bian:2016aa}
L.~Bian, J.~Suo, G.~Situ, \emph{et~al.}, \enquote{Multispectral imaging using a
  single bucket detector,} {{Scientific Reports}}
  \textbf{6}, 24752 (2016).

\bibitem{Pittman:1995}
T.~B. Pittman, Y.~H. Shih, D.~V. Strekalov, and A.~V. Sergienko,
  \enquote{Optical imaging by means of two-photon quantum entanglement,}
  {{Phys. Rev. A}} \textbf{52}, R3429--R3432 (1995).

\bibitem{Shih:24}
Y.~Shih, \enquote{Ghost imaging---its physics and application [invited],}
  {{Chin. Opt. Lett.}} \textbf{22}, 060011 (2024).

\bibitem{Chan:2008}
W.~L. Chan, K.~Charan, D.~Takhar, \emph{et~al.}, \enquote{A single-pixel
  terahertz imaging system based on compressed sensing,}
  {{Applied Physics Letters}} \textbf{93} (2008).

\bibitem{Candes:2008}
E.~J. Candes and M.~B. Wakin, \enquote{An introduction to compressive
  sampling,} {{IEEE Signal Processing Magazine}}
  \textbf{25}, 21--30 (2008).

\bibitem{Duarte:2008}
M.~F. Duarte, M.~A. Davenport, D.~Takhar, \emph{et~al.}, \enquote{Single-pixel
  imaging via compressive sampling,} {{IEEE Signal
  Processing Magazine}} \textbf{25}, 83--91 (2008).

\bibitem{Katz:2009}
O.~Katz, Y.~Bromberg, and Y.~Silberberg, \enquote{Compressive ghost imaging,}
  {{Applied Physics Letters}} \textbf{95}, 131110 (2009).

\bibitem{Gong:2012}
W.~Gong and S.~Han, \enquote{Experimental investigation of the quality of
  lensless super-resolution ghost imaging via sparsity constraints,}
  {{Physics Letters A}} \textbf{376}, 1519--1522 (2012).

\bibitem{Zhao:2012}
C.~Zhao, W.~Gong, M.~Chen, \emph{et~al.}, \enquote{Ghost imaging lidar via
  sparsity constraints,} {{Applied Physics Letters}}
  \textbf{101}, 141123 (2012).

\bibitem{Han:2018}
S.~Han, H.~Yu, X.~Shen, \emph{et~al.}, \enquote{A review of ghost imaging via
  sparsity constraints,} {{Applied Sciences}} \textbf{8}
  (2018).

\bibitem{Donoho:2006}
D.~L. Donoho, \enquote{Compressed sensing,} {{IEEE
  Transactions on information theory}} \textbf{52}, 1289--1306 (2006).

\bibitem{Tian:2022}
Y.~Tian, Y.~Fu, and J.~Zhang, \enquote{Plug-and-play algorithms for
  single-pixel imaging,} {{Optics and Lasers in
  Engineering}} \textbf{154}, 106970 (2022).

\bibitem{Lyu:2017aa}
M.~Lyu, W.~Wang, H.~Wang, \emph{et~al.}, \enquote{Deep-learning-based ghost
  imaging,} {{Scientific Reports}} \textbf{7}, 17865
  (2017).

\bibitem{He:2018aa}
Y.~He, G.~Wang, G.~Dong, \emph{et~al.}, \enquote{Ghost imaging based on deep
  learning,} {{Scientific Reports}} \textbf{8}, 6469
  (2018).

\bibitem{Wang:19}
F.~Wang, H.~Wang, H.~Wang, \emph{et~al.}, \enquote{Learning from simulation: An
  end-to-end deep-learning approach for computational ghost imaging,}
  {{Opt. Express}} \textbf{27}, 25560--25572 (2019).

\bibitem{Wu:20}
H.~Wu, R.~Wang, G.~Zhao, \emph{et~al.}, \enquote{Sub-nyquist computational
  ghost imaging with deep learning,} {{Opt. Express}}
  \textbf{28}, 3846--3853 (2020).

\bibitem{Yu:22}
Z.~Yu, Y.~Liu, J.~Li, \emph{et~al.}, \enquote{Color computational ghost imaging
  by deep learning based on simulation data training,}
  {{Appl. Opt.}} \textbf{61}, 1022--1029 (2022).

\bibitem{Hahamovich:2021aa}
E.~Hahamovich, S.~Monin, Y.~Hazan, and A.~Rosenthal, \enquote{Single pixel
  imaging at megahertz switching rates via cyclic {Hadamard} masks,}
  {{Nature Communications}} \textbf{12}, 4516 (2021).

\bibitem{Lin:25}
S.~Lin, Y.~Zhang, H.~Wang, \emph{et~al.}, \enquote{Learning-based high-speed
  single-pixel imaging using a cyclic random mask,} {{Opt.
  Express}} \textbf{33}, 25728--25742 (2025).

\bibitem{Kilcullen:2022aa}
P.~Kilcullen, T.~Ozaki, and J.~Liang, \enquote{Compressed ultrahigh-speed
  single-pixel imaging by swept aggregate patterns,}
  {{Nature Communications}} \textbf{13}, 7879 (2022).

\bibitem{Xu:18}
Z.-H. Xu, W.~Chen, J.~Penuelas, \emph{et~al.}, \enquote{1000 fps computational
  ghost imaging using led-based structured illumination,}
  {{Opt. Express}} \textbf{26}, 2427--2434 (2018).

\bibitem{Huang:2022}
H.~Huang, L.~Li, Y.~Ma, and M.~Sun, \enquote{25,000 fps computational ghost
  imaging with ultrafast structured illumination,}
  {{Electronic Materials}} \textbf{3}, 93--100 (2022).

\bibitem{Johnstone:24}
G.~E. Johnstone, J.~Gray, S.~Bennett, \emph{et~al.}, \enquote{High speed single
  pixel imaging using a microled-on-cmos light projector,}
  {{Opt. Express}} \textbf{32}, 24615--24628 (2024).

\bibitem{Zhao:2019}
W.~Zhao, H.~Chen, Y.~Yuan, \emph{et~al.}, \enquote{Ultrahigh-speed color
  imaging with single-pixel detectors at low light level,}
  {{Phys. Rev. Appl.}} \textbf{12}, 034049 (2019).

\bibitem{Stojek:2022}
R.~Stojek, A.~Pastuszczak, P.~Wr{\'o}bel, and R.~Koty{\'n}ski, \enquote{Single
  pixel imaging at high pixel resolutions,} {{Optics
  express}} \textbf{30}, 22730--22745 (2022).

\bibitem{Kanno:20}
H.~Kanno, H.~Mikami, and K.~Goda, \enquote{High-speed single-pixel imaging by
  frequency-time-division multiplexing,} {{Opt. Lett.}}
  \textbf{45}, 2339--2342 (2020).

\bibitem{Jiang:22}
W.~Jiang, Y.~Yin, J.~Jiao, \emph{et~al.}, \enquote{2,000,000~fps {2D} and {3D}
  imaging of periodic or reproducible scenes with single-pixel detectors,}
  {{Photon. Res.}} \textbf{10}, 2157--2164 (2022).

\bibitem{Hanawa:22}
J.~Hanawa, T.~Niiyama, Y.~Endo, and S.~Sunada, \enquote{Gigahertz-rate random
  speckle projection for high-speed single-pixel image classification,}
  {{Opt. Express}} \textbf{30}, 22911--22921 (2022).

\bibitem{Yamaguchi:2023aa}
T.~Yamaguchi, K.~Arai, T.~Niiyama, \emph{et~al.}, \enquote{Time-domain photonic
  image processor based on speckle projection and reservoir computing,}
  {{Communications Physics}} \textbf{6}, 250 (2023).

\bibitem{Wan:2025}
Y.~Wan, Z.~Long, X.~Fan, and Z.~He, \enquote{Scan-less speckle encoded
  single-pixel imaging over giga-pixels per second assisted by dual optical
  frequency combs,} {{Laser \& Photonics Reviews}} p.
  e01235.

\bibitem{Wang:2022aa}
F.~Wang, C.~Wang, M.~Chen, \emph{et~al.}, \enquote{Far-field super-resolution
  ghost imaging with a deep neural network constraint,}
  {{Light: Science \& Applications}} \textbf{11}, 1
  (2022).

\bibitem{Quan:2022}
Y.~Quan, X.~Qin, T.~Pang, and H.~Ji, \enquote{Dual-domain self-supervised
  learning and model adaption for deep compressive imaging,} in \emph{Computer
  Vision -- ECCV 2022: 17th European Conference, Tel Aviv, Israel, October
  23--27, 2022, Proceedings, Part XXX,}  (Springer-Verlag, Berlin, Heidelberg,
  2022), pp. 409--426.

\bibitem{Diamond2017Unrolled}
S.~Diamond, V.~Sitzmann, F.~Heide, and G.~Wetzstein, \enquote{Unrolled optimization with deep priors,} arXiv:1705.08041 (2017).

\bibitem{Liu:2019}
J.~Liu, Y.~Sun, X.~Xu, and U.~S. Kamilov, \enquote{Image restoration using
  total variation regularized deep image prior,} in \emph{ICASSP 2019-2019 IEEE
  International Conference on Acoustics, Speech and Signal Processing
  (ICASSP),}  (Ieee, 2019), pp. 7715--7719.

\bibitem{Rudin:1992}
L.~I. Rudin, S.~Osher, and E.~Fatemi, \enquote{Nonlinear total variation based
  noise removal algorithms,} {{Physica D: nonlinear
  phenomena}} \textbf{60}, 259--268 (1992).

\end{thebibliography}
%\begin{thebibliography}{99}

\end{document}

%\preprint{APS/123-QED}

\title{100 Mfps ghost imaging with wavelength division multiplexing: supplemental document}

\author{
Shin Motooka$^{1}$, 
Noriki Komori$^{1}$, 
Tomoaki Niiyama$^{2}$, 
and 
Satoshi Sunada$^{2}$
}

\affiliation{${}^{1}$Graduate School of Natural Science and Technology, Kanazawa University, Kakuma-machi, Kanazawa, Ishikawa, 920-1192, Japan\\ ${}^{2}$Faculty of Mechanical Engineering, Institute of Science and Engineering, Kanazawa University, Kakuma-machi, Kanazawa, Ishikawa, 920-1192, Japan}


\begin{abstract}

\end{abstract}

\maketitle
\section{Dynamic generation of multimode speckle patterns}
Let $u(t)$ denote a pseudorandom sequence generated by a waveform generator operating at 25~GS/s. The input laser light is phase-modulated with $u(t)$ and injected into a multimode fiber (MMF). The input electric field into an MMF is expressed as 
\begin{align}
\EE_{\rm in}(\rr,t) = \EE_{0}(\rr)e^{i(\alpha u(t)+\omega_0 t)},
\end{align}
where $\alpha$ represents the modulation strength, $\rr = (x,y)$ is the transverse coordinate, and $\omega_0$ is the angular frequency of the input light. 
The frequency component of the input field is given by
\begin{align}
\hat{\EE}_{\rm in}(\rr,\omega) 
&= \int \EE_{\rm in}(\rr,t)e^{-i\omega t}\,dt \nonumber \\
&= \left(
\int e^{i(\alpha u(t)+\omega_0 t)} e^{-i\omega t}\,dt
\right) \EE_0(\rr) 
\nonumber \\
&= \hat{U}(\omega)\EE_0(\rr),
\end{align}
where $\hat{U}(\omega)$ denotes the spectral component of the phase-modulated signal.

The electric field at position $z$ in the MMF can be expressed as \cite{Redding:13}, 
\begin{align}
\hat{\EE}(\rr,z,\omega) 
&= \sum_m^{M'} A_m(\omega) \PPsi_m(\rr,\omega)e^{-i\beta_m(\omega)z}, \\
 \EE(\rr,z,t) 
&= \int \hat{\EE}(\rr,z,\omega)e^{i\omega t}\,d\omega,
\end{align}
where $M'$ represents the number of guided modes excited by the input field, $A_m(\omega)$ is the complex amplitude of the $m$-th guided mode with spatial profile $\PPsi_m(\rr,\omega)$, and $\beta_m(\omega)$ is the propagation constant at angular frequency $\omega$. 
Assuming that $A_m(\omega)$ is simply determined by the boundary condition $\hat{\EE}(\rr,0,\omega) = \hat{\EE}_{in}(\rr,\omega)$, it can be written as
\begin{align}
A_m(\omega) = \hat{U}(\omega)\int \PPsi_m^*(\rr,\omega)\EE_0(\rr)\,d\rr = \hat{U}(\omega)a_m(\omega),
\end{align}
if the guided modes are orthogonal:
\begin{align}
\int \PPsi_m^*(\rr,\omega)\PPsi_{m'}(\rr,\omega)\,d\rr = \delta_{mm'}.
\end{align}

The output field at the end of an MMF with length $L$ at time $t$ is expressed as
\begin{align}
 \EE(\rr,L,t) 
&= \dfrac{1}{2\pi} 
\int \hat{\EE}(\rr,L,\omega)e^{i\omega t}\,d\omega \nonumber \\
&= \dfrac{1}{2\pi} \sum_m^{M'} \int A_m(\omega)\PPsi_m(\rr,\omega)e^{-i\beta_m(\omega)L} e^{i\omega t}\,d\omega \nonumber \\
&= \dfrac{1}{2\pi} \sum_m^{M'} \int a_m(\omega)\hat{U}(\omega)e^{-i\beta_m(\omega)L}e^{i\omega t}\PPsi_m(\rr,\omega)\,d\omega.
\end{align}

It is reasonable to assume that the spectrum of the modulation signal $u(t)$ is narrow, so that 
\begin{align}
\beta_m(\omega) = \beta_m(\omega_0) + \del{\beta_m(\omega_0)}{\omega}{}\Delta\omega + O\left(|\Delta\omega|^2\right),
\end{align}
and 
\begin{align}
\PPsi_m(\rr,\omega) \approx \PPsi_m(\rr,\omega_0), 
\hspace{1cm} a_m(\omega) \approx a_m(\omega_0)
\end{align}
where $\Delta\omega = \omega - \omega_0$. 
Consequently, the electric field can be approximately expressed as 
\begin{align}
 \EE(\rr,L,t) 
&\approx 
\dfrac{1}{2\pi} 
\left(
\sum_m^{M'} a_m(\omega_0)\PPsi_m(\rr,\omega_0)e^{-i\beta_m^0 L}
\int \hat{U}(\omega_0+\Delta\omega)e^{i\Delta\omega(t-\tau_m)} 
\,d\Delta\omega 
\right) e^{i\omega_0 t} \nonumber \\
%&= \sum_m^{M'} \tilde{a}_m(\omega_0)\PPsi_m(\rr,\omega_0)U(t-\tau_m)e^{i\omega_0 t} \nonumber \\
&= \sum_m^{M'} \tilde{a}_m(\omega_0)\PPsi_m(\rr,\omega_0)e^{i\alpha u(t-\tau_m)+i\omega_0 t},
\end{align}
where $\beta_m^0 = \beta_m(\omega_0)$ and $\tau_m = L\left.{\partial \beta_m(\omega)}/{\partial \omega}\right|_{\omega = \omega_0}$, and  
$\tilde{a}_m(\omega_0) =  a_m(\omega_0)e^{-i\beta_m^0 L}$. $\tau_m$ represents the group delay and depends on the mode index $m$ due to the MMF dispersion.
Accordingly, the MMF generates spatiotemporal speckle dynamics described by $I(\rr,t) = |\EE(\rr,L,t)|^2$ at the end facet, which can be used as the mask patterns for ghost imaging (GI). 

The MMF speckle pattern $I(\rr,t)$ strongly depends on the input wavelength $\lambda = 2\pi c/\omega_0$, where $c$ is the speed of light in vacuum. It has been reported that the correlation between speckle patterns generated by different wavelengths approaches zero when the wavelength difference satisfies $\Delta \lambda \ge \lambda^2/(n_{\mathrm{ref}} L \mathrm{NA}^2)$, where $n_{\mathrm{ref}}$ and $\mathrm{NA}$ are the refractive index and numerical aperture of the MMF, respectively \cite{Redding:13}. In our WDM-GI experiment, we set $\Delta \lambda \approx 0.8~\mathrm{nm}$ to meet this condition, ensuring that the speckle correlations for different input wavelengths are sufficiently low.

\begin{figure}[htbp]
  \centering
  \includegraphics[bb=0 0 926 383, width=.9\linewidth]{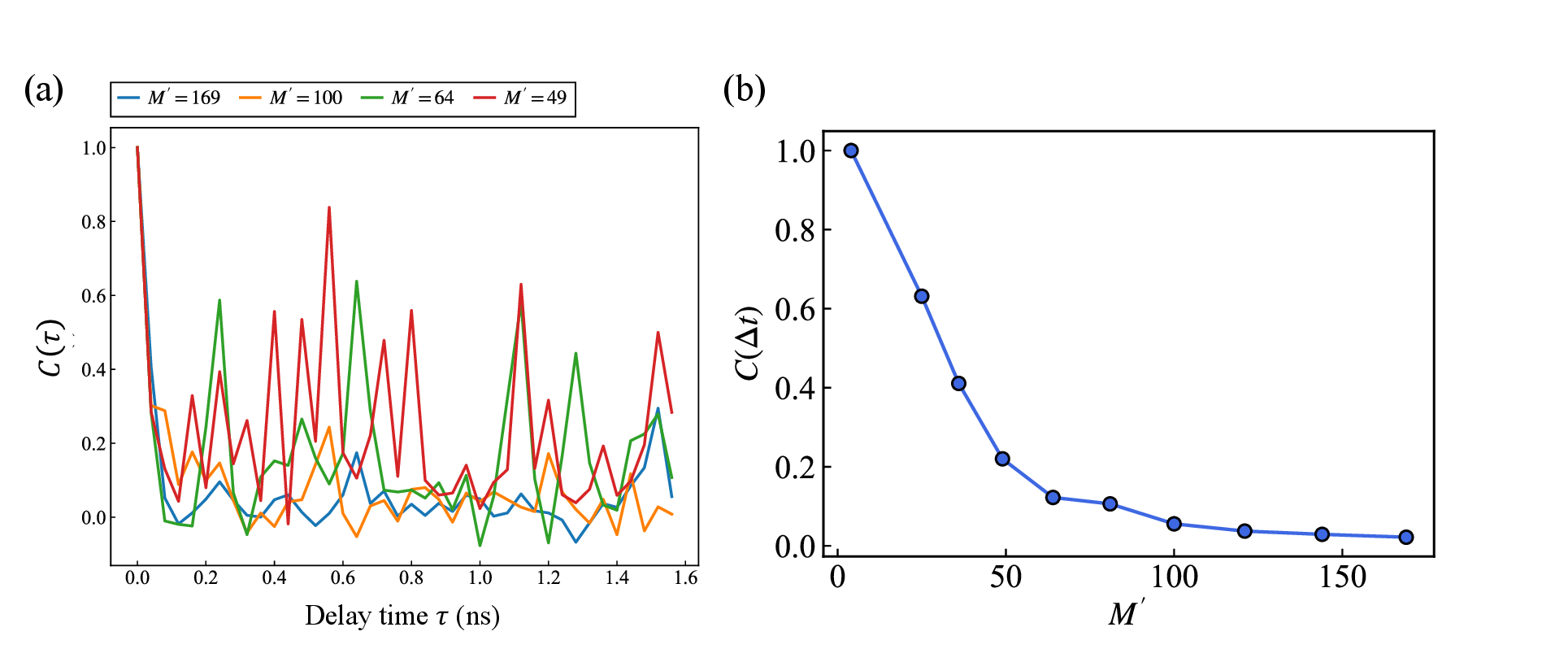}
  \caption{(a) $C(\tau)$ for $M'~=~$49, 64, 100, and 169. (b) $C(\Delta t)$ as a function of $M'$.}
\label{fig:spec}
\end{figure}

To evaluate the dynamical features of the generated speckle mask patterns, we measured the spatially averaged self-correlation function $C(\tau) = \langle I(\rr,0)I(\rr,\tau)\rangle_{\rr}/\sigma_I^2$, where $\sigma_I^2$ and $\langle\cdot\rangle_{\rr}$ denote the variance of $I(\rr,t)$ and the spatial average, respectively. 
Figure~\ref{fig:spec}(a) shows the correlation $C(\tau)$ as a function of the delay time $\tau$ for various $M'$ values under the injection of a random signal $u(t)$ that satisfies $\langle u(t_i)u(t_j)\rangle = \delta_{ij}$ with a sampling interval of $\Delta t = 0.04$~ns. We assumed that all the guided modes were excited with the complex amplitudes, $\{\tilde{a}_m\}_{m}^{M'}$, generated from a uniform distribution. The correlation rapidly decreases at $\tau = \Delta t = 0.04$~ns for $M' \ge 100$.  
The dependence of $C(\Delta t)$ on the number of guided modes $M'$ is shown in Fig.~\ref{fig:spec}(b). 

The maximum number of guided modes excited by the input field can be roughly estimated as $M' \approx V^2/2$, where $V = \pi a\,NA/\lambda$ denotes the V-number of the MMF with core diameter $a$ and numerical aperture $NA$ at the input wavelength $\lambda = 2\pi c/\omega_0$. 
In our experiment, we used an MMF with $a = 200~\mu$m and $NA = 0.39$ at $\lambda \approx 1550$~nm, suggesting a maximum mode number of $M' > 10^4$. 
However, the estimated $M'$ is based on the assumption that all modes are excited, whereas the actual number of excited modes may strongly depend on the coupling conditions of the input field.

\section{Experimental Setup}
The details of the experimental setup used in this study are as follows. 
A narrow-linewidth tunable laser (Alnair Labs, TLG-220, linewidth < 100~kHz, 30~mW) was employed as the light source, and five wavelengths were selected based on the ITU-T grid standard. 
Random waveforms were generated by a waveform generator (Tektronix, AWG70002A, 25~GS/s) and used for the phase modulation using a lithium niobate phase modulator (EO Space, PM-5S5-20-PFA-PFA-UV-UL, 16~GHz bandwidth).
The modulated light was transmitted through a polarization-maintaining single-mode fiber and introduced into a step-index multimode fiber (MMF, core diameter 200~$\mu$m, NA~0.39, length~20~m), where temporally varying speckle patterns were generated through the modal interference (see Sec.~1). 

The target images were displayed on a digital micromirror device (DMD, 800 $\times$ 1280, pixel pitch 10.6~$\mu$m) within a partial region of 200 $\times$ 200 pixels. 
The reflected light from the DMD was collected by a focusing lens, guided into an MMF with a core diameter of 105~$\mu$m, and delivered to a multimode photodetector (Newport 1544-B-50, DC-coupled InGaAs photoreceiver, 12~GHz bandwidth). 
The acquired signals were digitized and recorded using a digital oscilloscope (Tektronix, DPO72504DX, 25~GHz bandwidth) and transferred to a computer for reconstruction processing. 
To improve the signal-to-noise ratio (SNR), the target images and their inverted counterparts should be alternately displayed, and the differential signals between the two can be computed.

It is important to note that speckle patterns at different wavelengths are independent of each other.
In a practical WDM-GI system, independent detection channels are typically required to separately measure the optical signals corresponding to each wavelength.
However, for the purpose of verifying the feasibility of our approach, it is sufficient to use a single detection channel, as the measurements can be performed sequentially for each wavelength while maintaining the independence of the speckle patterns.
Therefore, in our proof-of-concept experiment, we used a single photoreceiver as the receiver unit for the WDM-GI.

\section{Comparison of Reconstruction Algorithms Using High-Resolution Images}
\begin{figure}[htbp]
  \centering
  \includegraphics[bb=0 0 360 355, width=.4\linewidth]{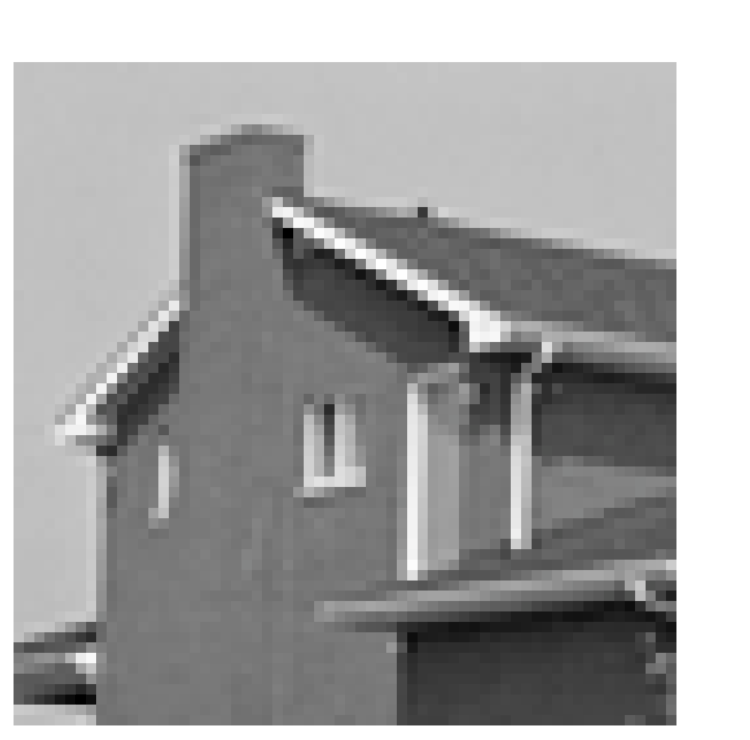}
  \caption{Target image of size 64$\times$64 pixels. The House image from the Set11 dataset \cite{Set11} was downsampled to a 64$\times$64 resolution in this study.}
\label{fig:target-image}
\end{figure}

\begin{figure}[htbp]
  \centering
  \includegraphics[bb=0 0 817 361, width=1\linewidth]{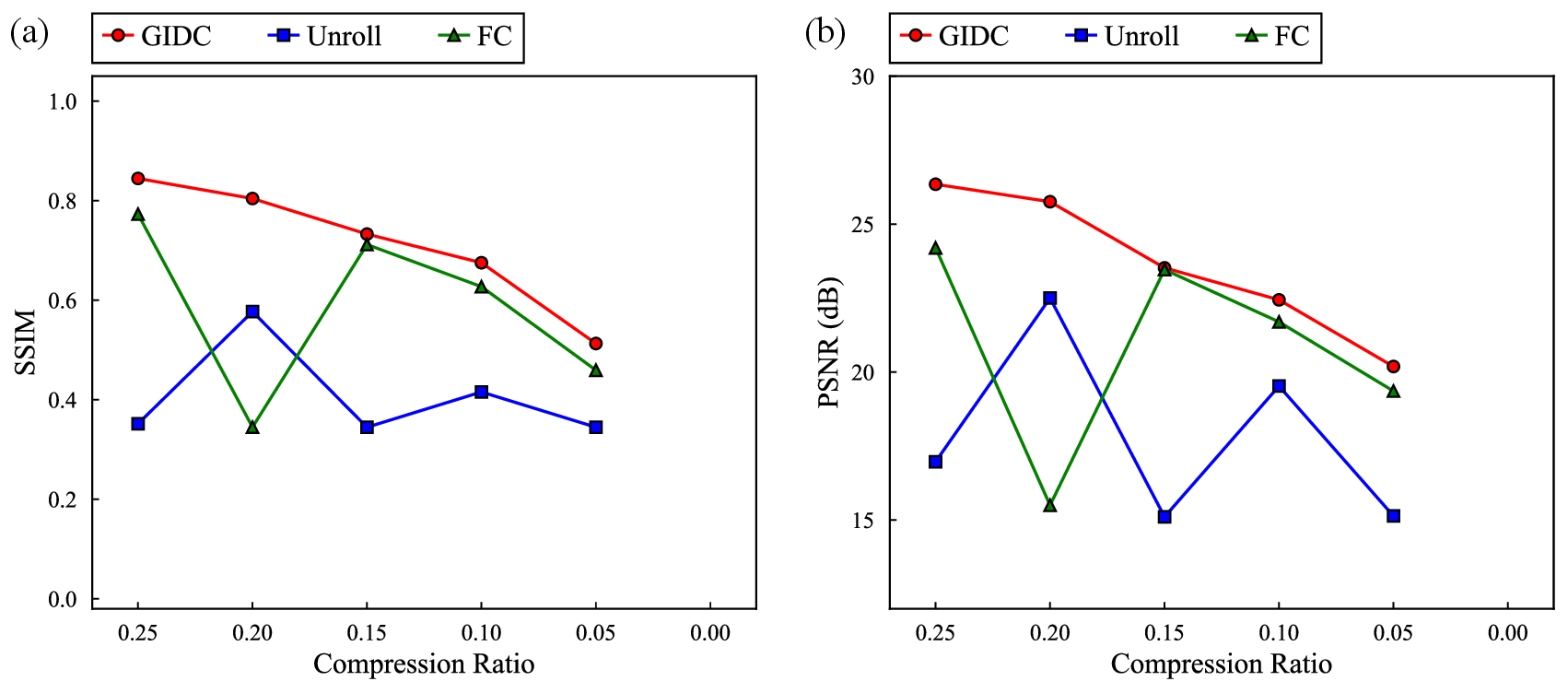}
  \caption{Performance comparison of reconstruction models (GIDC, Unrolling CNN, and FC model). (a) Structural Similarity Index Measure (SSIM). (b) Peak Signal-to-Noise Ratio (PSNR). $K = 1$.}
\label{fig:reconstruction-performance}
\end{figure}

\begin{figure}[htbp]
    \centering
    \includegraphics[bb=0 0 703 389,width=1\linewidth]{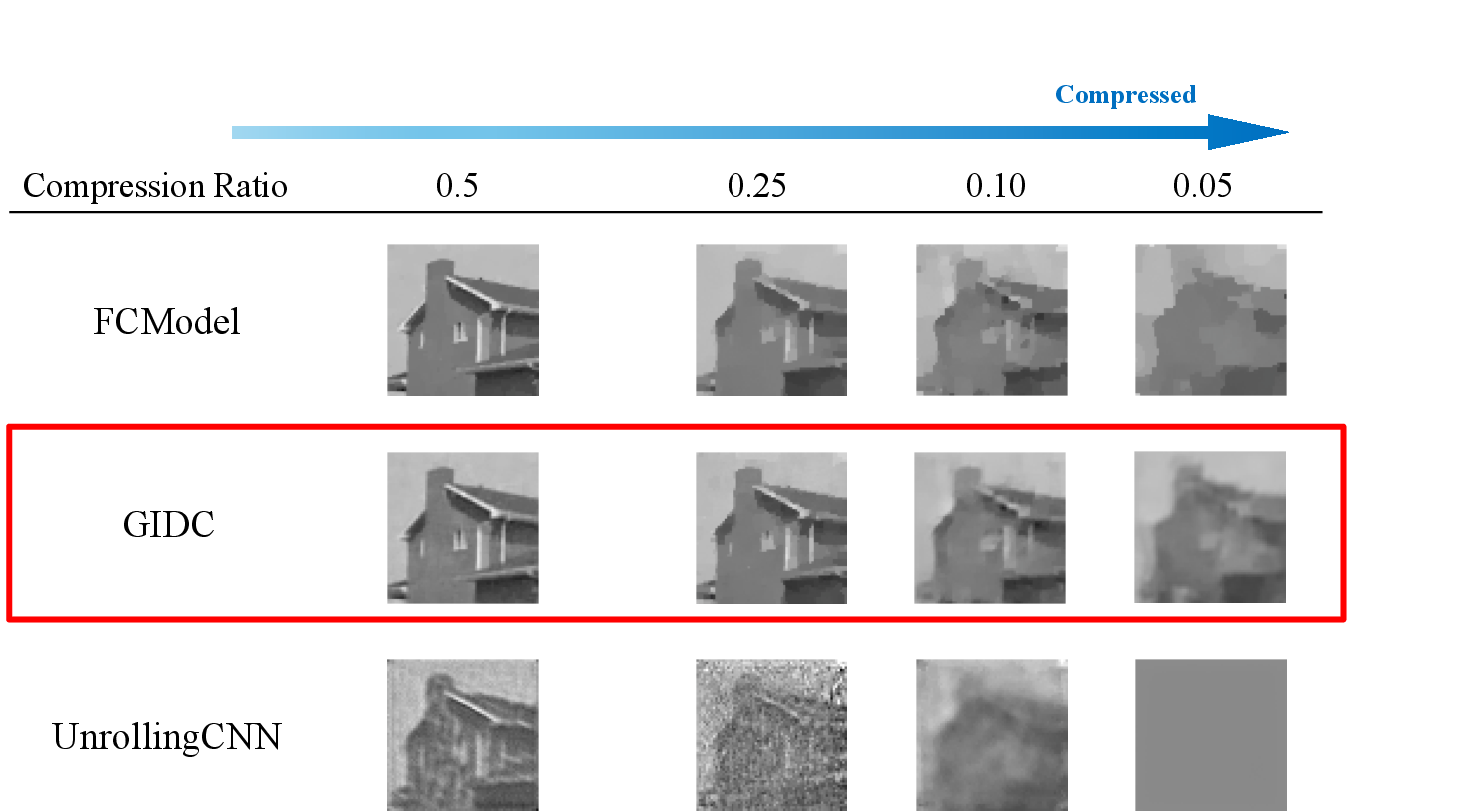}
    \caption{Reconstruction results of different models for the House image (64$\times$64 pixels). GIDC (highlighted in red) consistently achieves superior visual quality, particularly at low compression ratios. $K~=~1$.}
    \label{fig:comparison-images}
  \end{figure}

We numerically evaluated the performance of several reconstruction models for the target image shown in Fig.~\ref{fig:target-image}. The models considered in this study were GIDC, Unrolling CNN, and a fully connected (FC) model. The Unrolling CNN implements the proximal operator in the proximal gradient method through learnable CNN parameters--a technique widely adopted in recent studies on compressive sensing~\cite{quan2022dual,Diamond2017Unrolled}. The FC model employs a simple architecture consisting of a single fully connected hidden layer. Figures~\ref{fig:reconstruction-performance} and \ref{fig:comparison-images} show the reconstruction results obtained for a single wavelength (1550.12~nm). The compression ratio was defined as $N/(KM)$, where $N$, $M$, and $K$ represent the numbers of pixels, mask patterns, and input wavelengths, respectively. As shown in these figures, GIDC outperforms both the Unrolling CNN and FC model, maintaining high reconstruction quality even at low compression ratios.

\section{Gaussian vs. Speckle Masks}
\begin{figure}[htbp]
  \centering
  \includegraphics[bb=0 0 888 354,width=0.9\linewidth]{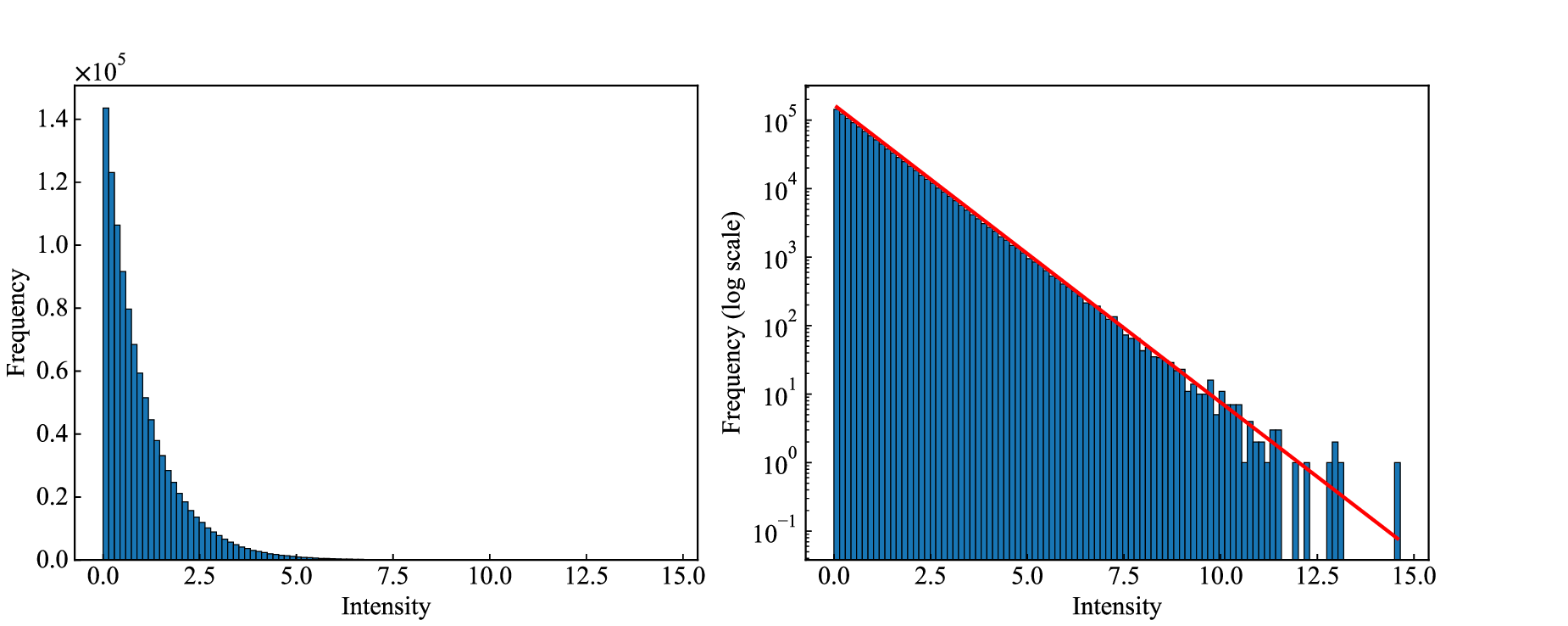}
  \caption{Distribution of speckle pattern intensity used in this study. The blue bars represent the measured data, and the red curve indicates the exponential fit $e^{-\lambda I}$ with parameter $\lambda = 1.0$.}
  \label{fig:distribution-comparison}
\end{figure}

\begin{figure}[htbp]
  \centering
  \includegraphics[bb=0 0 716 533,width=1\linewidth]{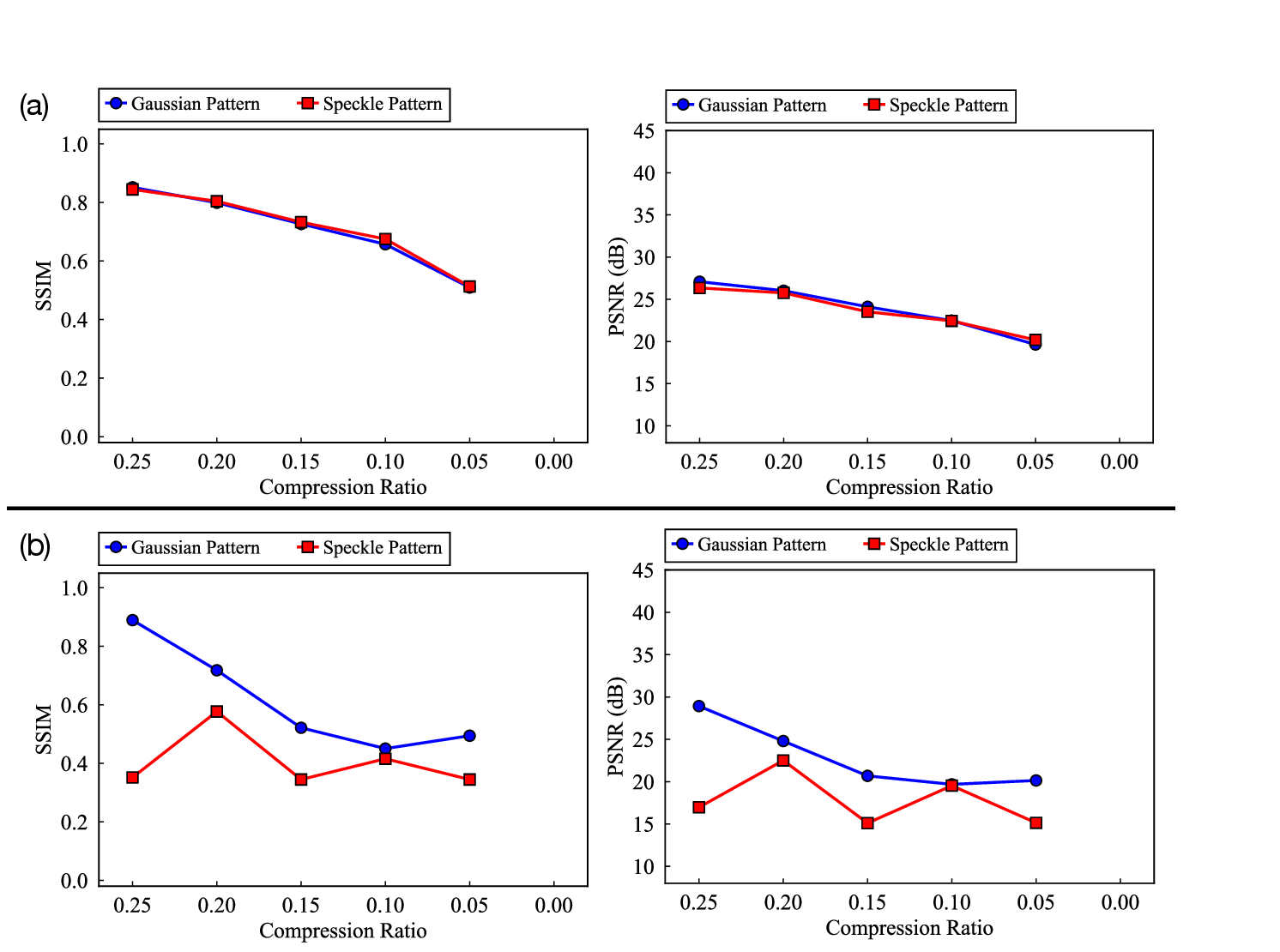}
  \caption{Comparison of reconstruction performance for speckle mask patterns (exponential distribution) and Gaussian mask patterns. (a) GIDC achieves consistent SSIM and PSNR regardless of the mask distribution. (b) Unrolling CNN exhibits substantial performance variation depending on the distribution.}
  \label{fig:gaussian-vs-exponential}
\end{figure}

Here, we numerically evaluate the reconstruction performance using two types of mask patterns. The first is a Gaussian mask, widely adopted in compressed sensing. The second is a speckle mask, characterized by an exponentially distributed intensity (Fig.~\ref{fig:distribution-comparison}). The comparison results are shown in Fig.~\ref{fig:gaussian-vs-exponential}. In this experiment, we used the $64 \times 64$ pixels \textit{House} image from the Set11 dataset~\cite{Set11} as the target. 

A comparison between Figs.~\ref{fig:gaussian-vs-exponential}(a) and \ref{fig:gaussian-vs-exponential}(b) indicates that the reconstruction performance of GIDC (Fig.~\ref{fig:gaussian-vs-exponential}(a)) is largely insensitive to the mask pattern type, whereas the Unrolling CNN exhibits pronounced dependence on it. This robustness of GIDC motivated its adoption as the reconstruction model in this study.

% \begin{condenseditemize}
% \item[] Algorithm S1
% \item[] Equation (S1)
% \item[] Figure S1
% \item[] Media S1
% \item[] Table S1
% \end{condenseditemize}

\section{Measurement Error of Speckle Patterns}
\begin{figure}[htbp]
  \centering
  \includegraphics[bb=0 0 580 504,width=1\linewidth]{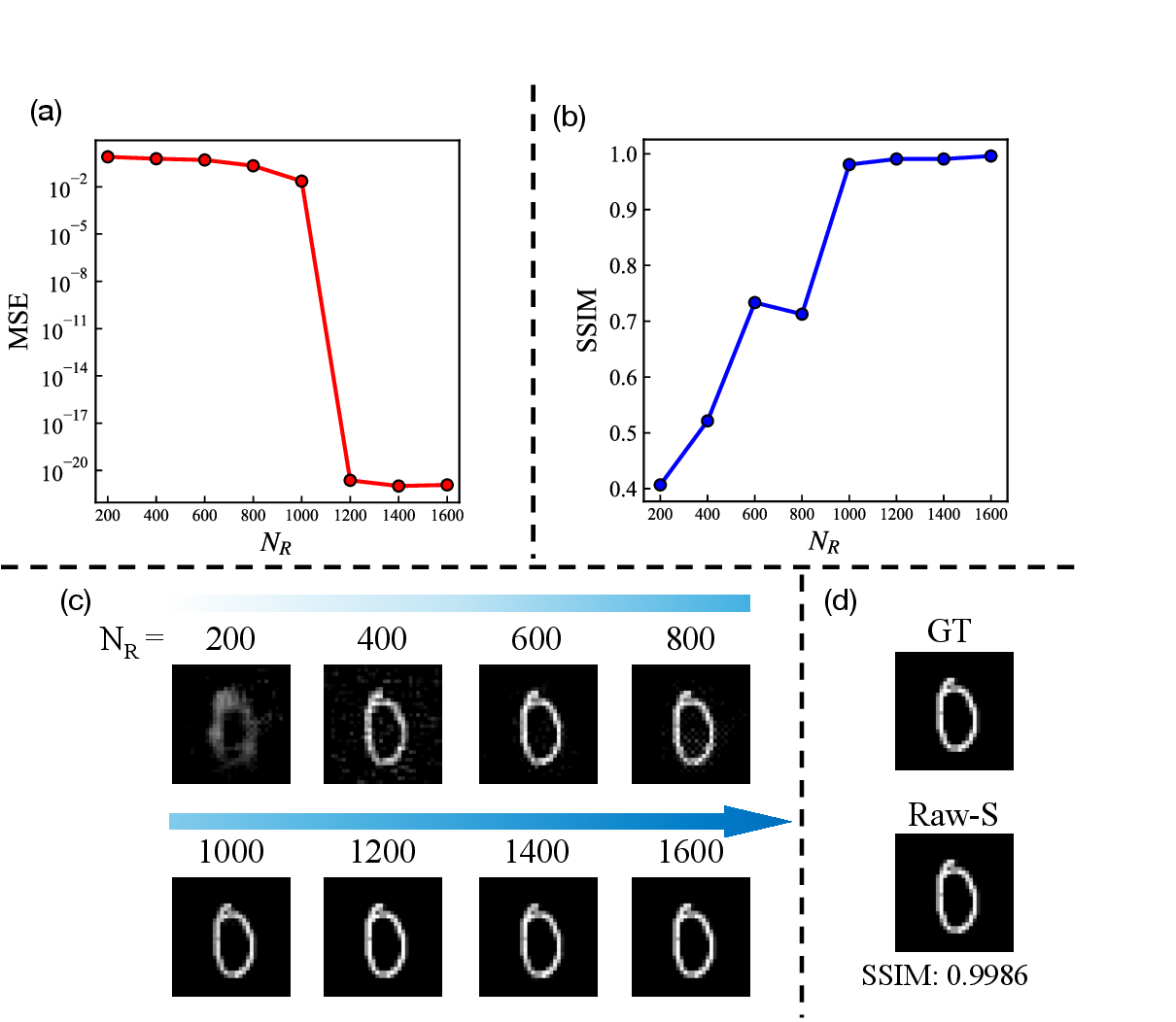}
  \caption{
  (a) MSE as a function of $N_R$. 
  (b) SSIM of reconstructed images as a function of $N_R$. 
  (c) Reconstructed (digit ``0'') images for $N_R =$ 200, 400, 600, 800, 1000, 1200, 1400, and 1600. 
  (d) Ground truth (GT) image and reconstructed image obtained using the ground-truth measurement matrix $\Ss$ (Raw-S).}
  \label{fig:est-mask}
\end{figure}

Dynamic speckle patterns were measured using multiple a priori images $\xx_r$ $(r \in {1, 2, \cdots, N_R})$. To investigate how the measurement error depends on the number of images $N_R$, we performed numerical simulations.
Figure~\ref{fig:est-mask}(a) shows the mean squared error (MSE) between the estimated and ground-truth speckle patterns as a function of $N_R$, where the speckle image resolution was set to $N = 32 \times 32$.
Figure~\ref{fig:est-mask}(b) presents the structural similarity index measure (SSIM) of the reconstructed images obtained using the estimated speckle mask patterns.
As shown in Fig.~\ref{fig:est-mask}(a), the MSE decreases markedly when $N_R > N$, leading to improved reconstruction quality. Consequently, the reconstructed images more closely resemble the ground truth (GT), particularly for $N_R > N$, as illustrated in Fig.~\ref{fig:est-mask}(b).
Representative reconstructed images are shown in Fig.~\ref{fig:est-mask}(c).
For comparison, Fig.~\ref{fig:est-mask}(d) displays the GT and the reconstructed images obtained using the ground-truth measurement matrix $\Ss$ (Raw-S).

\section{Achievable Imaging Speeds in WDM-GI under Supervised Learning}
We employed a self-supervised learning-based reconstruction approach that does not rely on ground-truth images. This approach enables imaging at frame rates of up to 100 Mfps, making it particularly suitable for scenarios where training data are unavailable or when previously unobserved phenomena must be captured.

\begin{figure}[htbp]
  \centering
  \includegraphics[bb=0 0 575 511,width=1\linewidth]{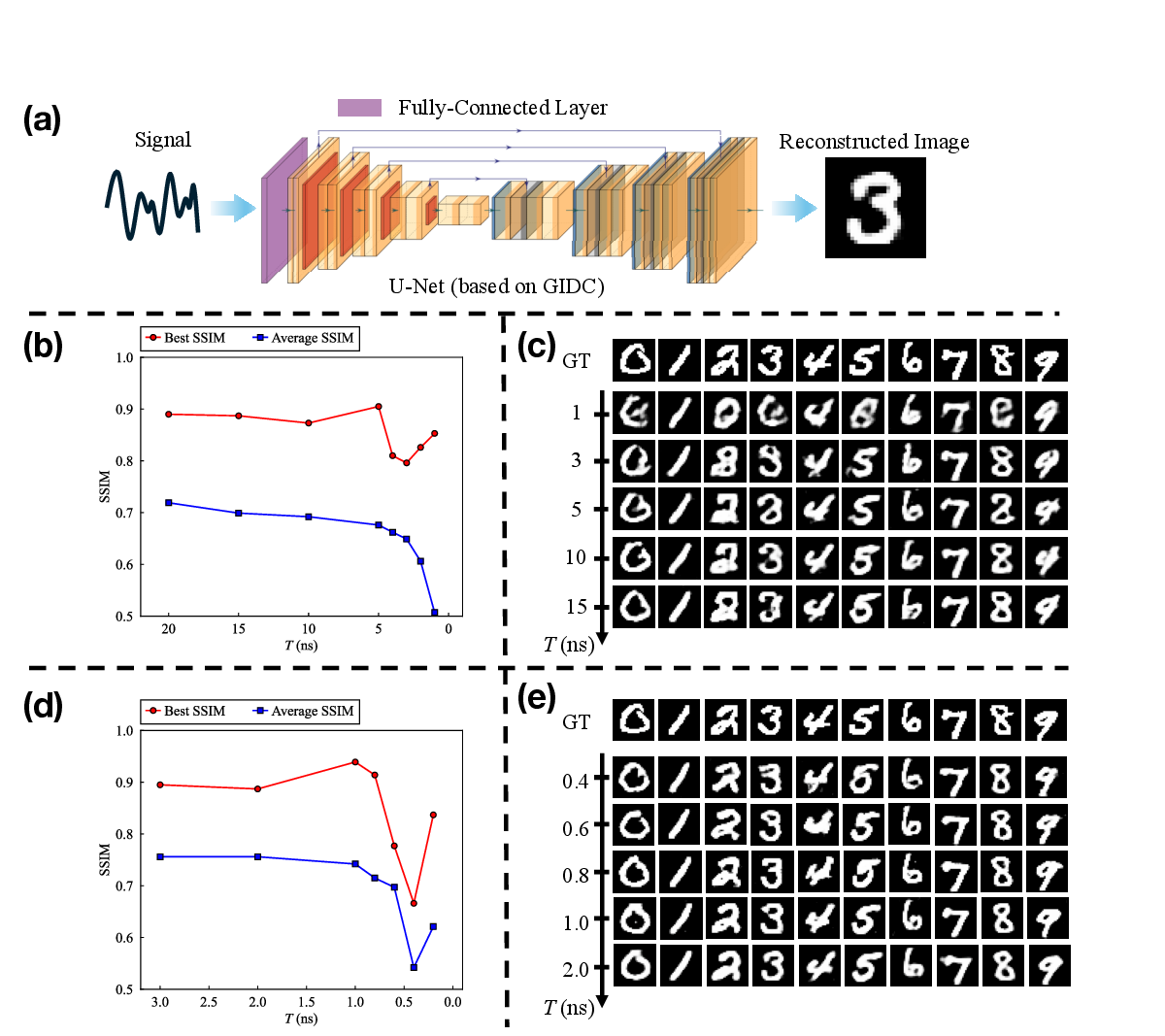}
  \caption{Supervised learning-based image reconstruction. (a) The reconstruction model consists of one FC layer followed by a U-Net. (b) SSIM as function of exposure time $T$ for a single wavelength ($K~=~1$).(c) Reconstructed images for $K~=~1$. (d) SSIM as a function of $T$ for $K~=~5$. (e) Reconstructed images for $K~=~5$.}
  \label{fig:supervised-learning}
\end{figure}

Here, we further investigated the performance limit achievable when ground-truth images are available and supervised learning can be applied. Using 2,000 MNIST images (1,900 for training, 90 for validation, and 10 for testing), we trained a reconstruction model consisting of a fully connected (FC) layer followed by a U-Net, as illustrated in Fig.~\ref{fig:supervised-learning}(a). The loss function was defined as the sum of the MSE between the target and reconstructed images and the total variation (TV) of the reconstructions.

Figure~\ref{fig:supervised-learning}(b) shows the SSIM for a single wavelength ($K=1$). The SSIM averaged over the 10 test images saturated for $T \ge 5$ ns. At shorter exposure times, slight artifacts appeared in the reconstructed images [Fig.~\ref{fig:supervised-learning}(c)], leading to a decrease in SSIM.

When multiple wavelengths were used ($K=5$), the averaged SSIM remained around 0.7 even at short exposure times of $T \approx 1$ ns [Fig.~\ref{fig:supervised-learning}(d)], and the reconstruction quality improved markedly [Fig.~\ref{fig:supervised-learning}(e)]. These results indicate that, when supervised learning is feasible, the proposed approach can achieve imaging at frame rates up to 1 Gfps, which is an order of magnitude higher than the already remarkable 100 Mfps attainable under the self-supervised setting.

\section{Impact of Quantization on Image Reconstruction}
\begin{figure}[htbp]
  \centering
  \includegraphics[bb=0 0 452 193,width=.6\linewidth]{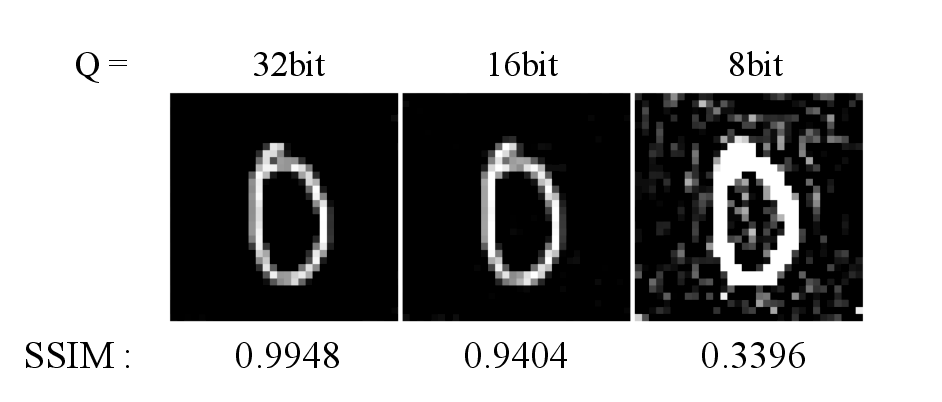}
  \caption{Effect of quantization on image reconstruction. From left to right: reconstruction results with 32-bit, 16-bit, and 8-bit quantization during AD conversion. Severe degradation is observed at lower bit depths, highlighting the sensitivity of reconstruction accuracy to quantization.}
  \label{fig:quan_img}
\end{figure}

Here, we assume that the acquired signals are quantized to $Q$ bits during the oscilloscope’s analog-to-digital (AD) conversion. To evaluate the influence of $Q$-bit quantization on the signals, we conducted numerical simulations. Figure \ref{fig:quan_img} shows the reconstruction results and the corresponding SSIM values for $Q$ = 32, 16, and 8. As expected, lower quantization levels degrade the quality of the reconstructed image.

\section{Demultiplexing}
\begin{figure}[htbp]
  \centering
  \includegraphics[bb=0 0 565 369, width=.7\linewidth]{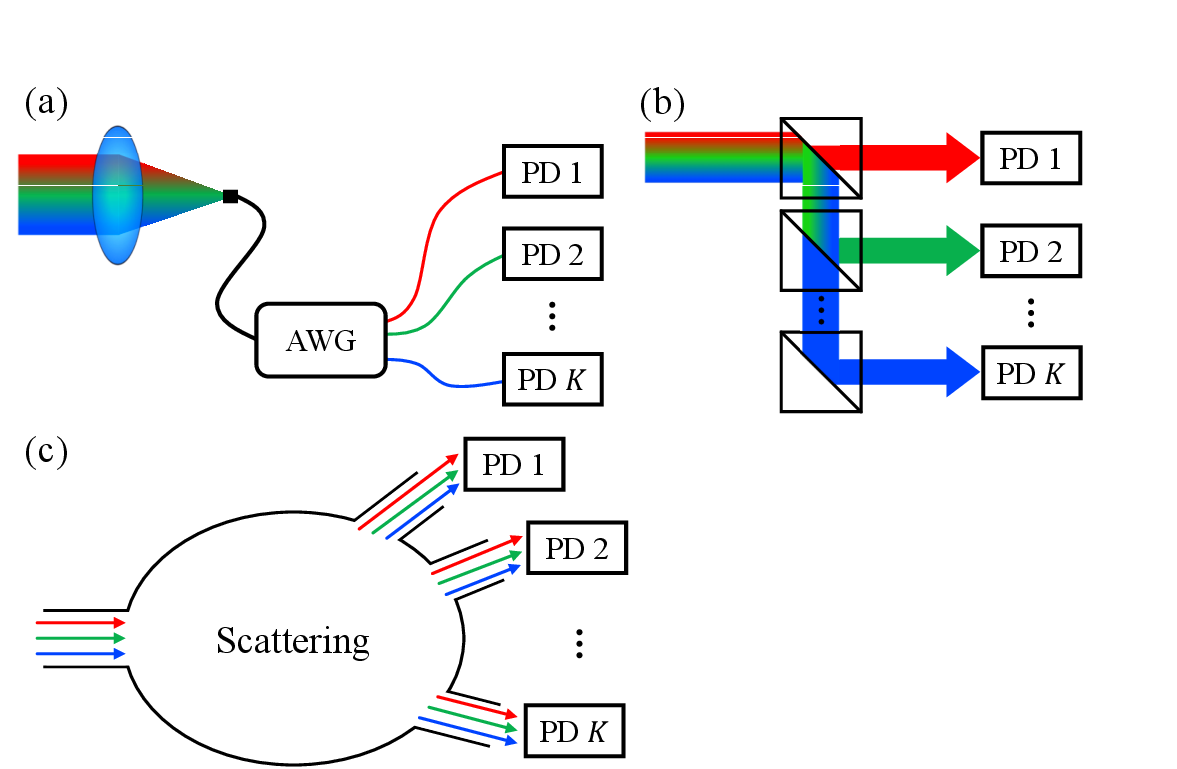}
  \caption{Demultiplexing methods using (a) arrayed waveguide grating (AWG), (b) space-optic system, and (c) optical scattering device.}
\label{fig_demux}
\end{figure}

Demultiplexing at the receiver unit can be implemented using various methods, as illustrated in Fig.~\ref{fig_demux}.
In particular, the approach shown in Fig.~\ref{fig_demux}(a) can be directly applied to the aforementioned fiber-based receiver unit.

\newpage
% Bibliography

%\bibliography{OPTICA_ref}
%\bibliographystyle{apsrev}
%\bibliography{apsrev.bst}
%\bibliography{/Users/sunada/Dropbox/My_Refs_db}
%\bibliography{ms_dfa-adjoint_ver9.bbl}
%\begin{thebibliography}{99}